\documentclass{article}
\usepackage{amsmath, amsthm, amsfonts}
\usepackage[margin=1.25in]{geometry}
\usepackage{amsmath}
\usepackage{amssymb}
\usepackage{IEEEtrantools}
\usepackage{framed}
\usepackage[small]{titlesec}
\usepackage{fancyhdr}
\usepackage{cases}
\usepackage{xcolor}

\usepackage{graphicx}
\usepackage{caption}
\usepackage{subcaption}
\usepackage{dblfloatfix}
\usepackage{float}

\usepackage[numbers,sort&compress]{natbib}
\usepackage{color}   
\usepackage{hyperref}
\hypersetup{
    colorlinks=true, 
    linktoc=all,     
    linkcolor=blue,  
}

\renewcommand\thesection{\Roman{section}}
\renewcommand\thesubsection{\Alph{subsection}}

\titleformat{\section}	
{\normalfont\bfseries\fillast}
{\bfseries \small {\thesection}.}
{1ex minus -10ex}
{\small}
\titlespacing{\section}
{3pc}{*3}{*2}[3pc]

\titleformat{\subsection}
{\normalfont\bfseries\fillast}
{\bfseries \small {\thesubsection}.}
{1ex minus -10ex}
{\small}
\titlespacing{\section}
{3pc}{*3}{*2}[3pc]

\titleformat{\subsubsection}
{\normalfont\bfseries\fillast}
{\bfseries \small {\thesubsubsection}}
{1ex minus -10ex}
{\small}
\titlespacing{\section}
{3pc}{*3}{*2}[3pc]

\newcommand{\D}[3][\text{}]{\frac{\partial ^#1 #3}{\partial #2^#1}}
\newcommand{\Do}[3][\text{}]{\frac{d ^#1 #3}{d #2^#1}}

\theoremstyle{definition}

\theoremstyle{remark}

\title{Results}
\date{\vspace{-5ex}}

\begin{document}

\pagestyle{myheadings}
\title{A Numerical Bifurcation Analysis of a Dryland Vegetation Model}
\author{C.B. Ward, P.G. Kevrekidis and N. Whitaker\\
Department of Mathematics and Statistics, University of Massachusetts\\
Amherst, MA 01003, USA
}
\maketitle

\begin{abstract}
  The dryland vegetation model proposed by Rietkerk and collaborators has been explored from a bifurcation perspective in several previous studies.
  Our aim here is to explore in some detail the bifurcation phenomena present
  when the coefficients of the model are allowed to vary in a wide range of
  parameters. In addition to the primary bifurcation parameter, the precipitation, we allow the two infiltration rate parameters to vary as well. We find that these two parameters control the size and stability of nonhomogeneous biomass states in a way that can be predicted. Further, they control when certain homogeneous and inhomogeneous (in space) periodic (in time) orbits exist. Finally, we show that the model possesses infinitely many unphysical steady state branches. We then present a modification of the model which eliminates these unphysical solutions, and briefly explore this new model for a fixed set of parameters.
\end{abstract}

\section{Introduction}\label{sec1}
Spontaneous pattern formation in semi-arid environments has been of increasing interest in the past two decades, with many systems being described by increasingly sophisticated PDE models \cite{16,17,18,19,20}. Much work has been done in the way of determining how these systems evolve in time and what changes occur due to changes in average precipitation; special emphasis has been placed on finding heuristic diagnostics for predicting ecosystem health \cite{21,22,23,24,25,26,27,28,29,4}.
One of the key drivers of this effort has been the consideration of
the robustness of spatial patterns and their sequences of bifurcations
in such systems~\cite{17}.

Inspired by the the fundamental work on morphogenesis in
mathematical biology~\cite{5,6}, some of the prototypical models
in this area rely on nonlinear partial differential equations
of the reaction-diffusion type. Simpler models consider
two-component settings where the quantities characterized
are the plant biomass and the water concentration~\cite{17}. Subsequently,
more detailed models have explored variations of this,
considering that the water can be partitioned into soil
water and surface water, only the former of which
contributes to the growth of the plant biomass~\cite{19,26}.
The dynamics of this latter class of models has now been
studied in some detail, especially in a two-dimensional
setting, with an eye towards exploring transitions between
patterned states, such as gaps, labyrinthine patterns and spots.

On the other hand, relatively little has been done in determining how these
systems are effected by variations in parameters other than the
average precipitation (such as, e.g., the infiltration rates which
characterize the conversion of surface into soil water), especially
from a bifurcation perspective. This, as well as an exploration
of analytical features (including some ``pathological'' ones, such
as the existence of unphysical steady states) are among the principal
features of the present study. In addition to identifying the
steady states and periodic orbits of the system and the associated
spectral stability and nonlinear dynamics,
we also briefly propose a variant of the model eliminating unphysical
solutions and discuss its bifurcation characteristics. Our presentation
is structured as follows. Firstly, in section II, we present the model and
discuss its main features. Then, in section III, we perform a two-parameter exploration of the
model's characteristics, for different domain sizes (i.e., the role
of the domain size is considered). In section IV, we then identify the unphysical
branches and their implications, and in section V, we amend the original model so as to avoid
these unphysical branches. Finally, in section VI, we summarize our findings and discuss some
possibilities for future studies. In the appendix, some of
the analytical results (e.g., the Turing analysis) and
numerical methods are further discussed.

\section{Theoretical Model and Analytical Considerations}

In the present study, we will focus on the model by Rietkerk \emph{et al.}~\cite{19} --which for notational simplicity will be denoted as ``RM'',
but use the nondimensional form given by Zelnik \emph{et al.}~\cite{26}:
\begin{align}\label{eq1}
\D{t}{n} &= -\mu n + \frac{w}{w+1} n + \D[2]{x}{n} \nonumber \\ 
\D{t}{w} &= -\nu w + \alpha \frac{n + f}{n +1} h - \gamma \frac{w}{w+1} n + D_w \D[2]{x}{w} \\ 
\D{t}{h} &= p - \alpha \frac{n + f}{n +1} h + D_h \D[2]{x}{h} \nonumber
\end{align}
where $n$ is the plant biomass, $w$ is the soil water, and $h$ is the surface water. 

Each term in the above set of equations models a different physical process. Looking at the equation for the biomass, the first term determines the biomass decay
rate, the second term the biomass growth rate (due to the presence of
soil water $w$), and the third the biomass dispersal. The growth rate for biomass
through the $\frac{w}{w+1}$ term contains a saturation effect.
If the soil water concentration is small, then the growth rate is essentially linear in both biomass and soil water. On the other hand,
if the soil water is large, then the growth rate loses its dependence on soil
water and grows at a rate proportional solely to the biomass concentration. 

The second PDE governs the spatio-temporal evolution of the
soil water concentration. The first term is the rate at which soil water evaporates out of the soil, and is assumed to depend linearly on the soil water.
The second is the rate at which surface water is turned into soil water, i.e.,
the infiltration rate. Similarly to the growth term for biomass,
here too, there exists a saturation term. When the biomass is large,
the infiltration rate approaches a maximum and becomes independent of biomass. Similarly, when the biomass is small, the infiltration rate approaches
a minimum and effectively becomes independent of the biomass. This implies that in the realm of the
RM, surface water is converted to soil water at a monotonically,
with respect to the biomass, faster rate.
The third term is the rate at which the biomass is absorbing
the soil water. The term is identical to that of the biomass growth term except it is multiplied by a (small) constant prefactor $\gamma$. The fourth term
reflects the dispersal of the soil water.

The RM model was orginally based on that of Klausmeier~\cite{17}, a two species model. In going from two species to three, they distinguished between water in the soil and a small film of water on the soil's surface. The third PDE above was added to model this new surface water term, in line with what is expected in realistic settings of dryland
vegetation.
In the dynamic evolution equation for the surface water,
the first (constant) growth rate represents the annual average precipitation,
assumed to vary slowly compared to biomass, soil water, and surface water. The second and third terms are the infiltration and dispersal rates. Because the surface water is free to move unhindered by the soil, it is natural to expect
that $D_h \gg D_w$.

Examining the above set of equations from a dynamical systems viewpoint,
we immediately notice that it possesses
two homogeneous steady states. The first solution is a desert state given by $(n_1,w_1,h_1) = (0,\frac{p}{\nu},\frac{p}{\alpha f})$, while the second solution, a ``vegetated''  state (i.e., one with non-vanishing biomass), is given by
\begin{eqnarray}
  n_2 = \frac{1}{\gamma \mu} (p - \frac{\nu \mu}{1-\mu}) \quad , \quad w_2 = \frac{\mu}{1-\mu} \quad, \quad h_2 = \frac{p}{\alpha} \frac{n_2 + 1}{n_2 +f}
  \label{stst}
  \end{eqnarray}
The two solutions intersect at a {\it transcritical} bifurcation when $p=\frac{\mu \nu}{1-\mu}$. The desert state is stable to all perturbations for $p<\frac{\mu \nu}{1-\mu}$ and unstable to homogeneous perturbations for $p>\frac{\mu \nu}{1-\mu}$. Such perturbations in this case lead to the vegetated state.
Likewise, near the bifurcation point, the vegetated state is unstable to homogeneous perturbations for $p<\frac{\mu \nu}{1-\mu}$ and stable to all perturbations for $p>\frac{\mu \nu}{1-\mu}$.
In summary, at this point there is an exchange of stability between
the desert state (stable before this critical parameter) and the vegetated
state (stable past the critical point).
It is relevant to note that the desert state becomes unphysical (one of the species becomes negative) when $p<0$ and the vegetated state  becomes unphysical when $p<\frac{\mu \nu}{1-\mu}$ (the biomass concentration becomes
negative). Previous studies have essentially ignored these regions. We will see in the third section of this paper that these two unphysical solution branches are not the only ones.

\begin{figure}
\centering
        \includegraphics[width=.6\textwidth, keepaspectratio=true]{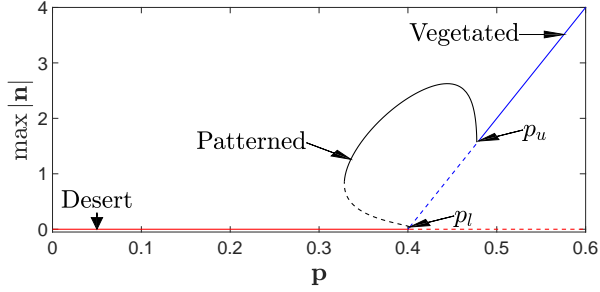}
        \caption{Bifurcation diagram of the RM on a small domain,
          showing the maximal biomass as a function of
          the precipitation parameter $p$. Here $\alpha=0.4$ and
          $f=0.2$.
          Note that we have not shown the unphysical portion of the vegetated branch (which
          is relevant in order to have a complete sense of the
          transcritical bifurcation
          at $p=\frac{\mu \nu}{1-\mu}$). The transcritical bifurcation renders
          the vegetated branch stable. However, the Turing (symmetry
          breaking) bifurcation
          at $p_l$ occurs soon after, making the branch unstable towards
          spatially inhomogeneous steady states.}
        \label{fig:Fig1}
\end{figure}

A Turing analysis~\cite{5,6} on the vegetated state shows that it becomes unstable to \emph{inhomogeneous}  perturbations between two $p$-values greater than $\frac{\mu \nu}{1-\mu}$, say on the interval $(p_l,p_u)$ \cite{4}. Between these two points, a number of bifurcations can also occur depending on the size of the domain, each giving rise to spatially inhomogeneous steady states, i.e. a patterned state~\cite{26}. Fig.~(\ref{fig:Fig1}) shows a bifurcation diagram of the desert state, the vegetated state, and a patterned state; see the  appendix for a description of the numerical methods used in this paper.

Our principal aim in what follows is to explore
the effects caused by changes to the infiltration rate term, as well
as the precipitation parameter.
To that end, we fix the following parameters at the value given by \cite{26}\\
\[\mu = 0.5 \quad \nu=0.4 \quad \gamma=0.1 \quad D_w=1 \quad D_h=1000\]
and allow the parameters $\alpha$, and $f$ which are related to the infiltration rate to be varied in individual sections. The primary bifurcation parameter is the precipitation $p$, which is allowed to vary from $0 \leq p \leq 0.6$ (except in section III where we extend this range). 

\section{Two Parameter Numerical Exploration}

\subsection{Small Domain Size}
\begin{figure}
\centering
\begin{subfigure}[c]{0.45\textwidth}
        \includegraphics[width=\textwidth, keepaspectratio=true]{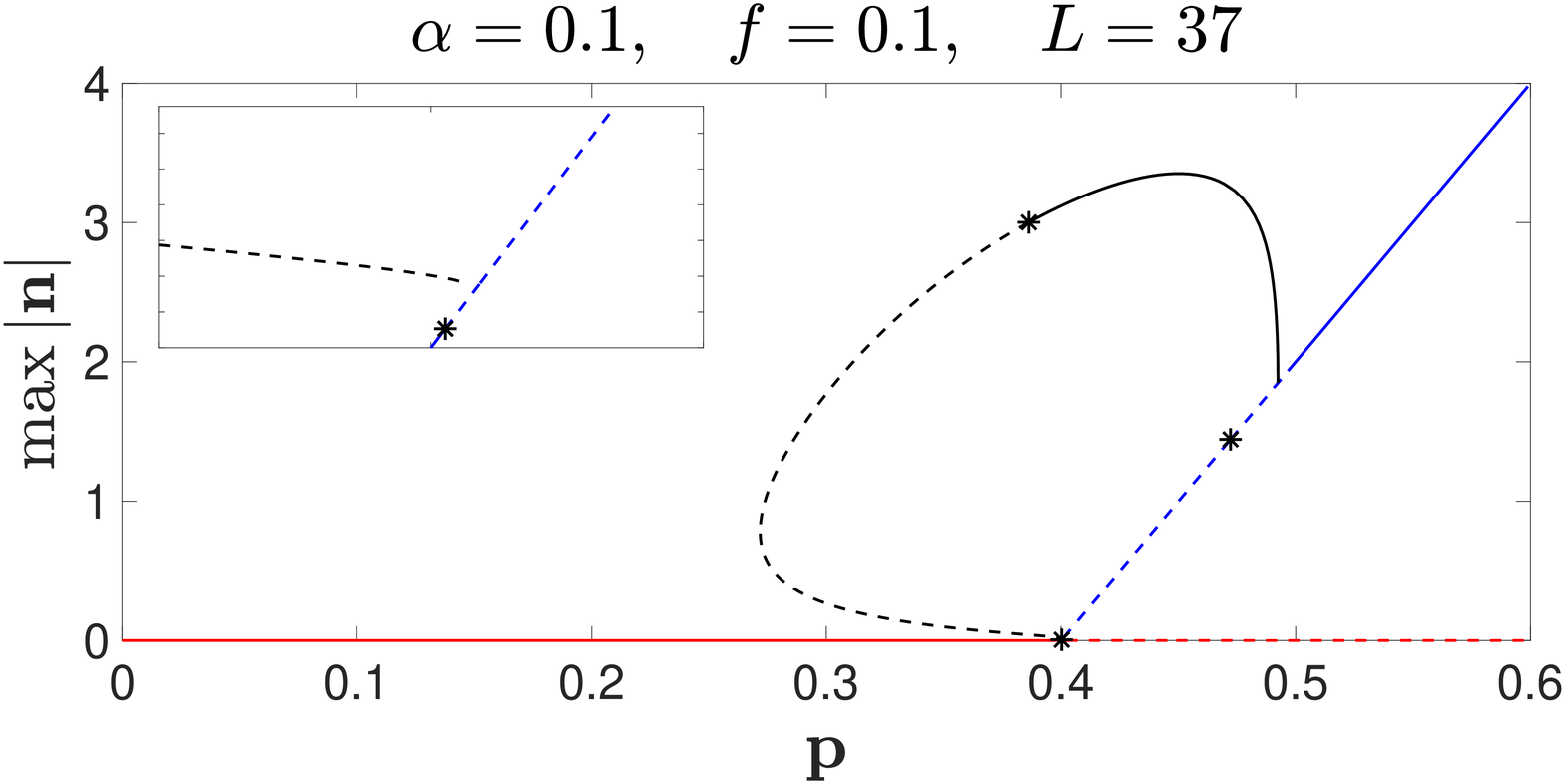}
        \caption*{(a)}
    \end{subfigure}
    \qquad
\begin{subfigure}[c]{0.45\textwidth}
        \includegraphics[width=\textwidth, keepaspectratio=true]{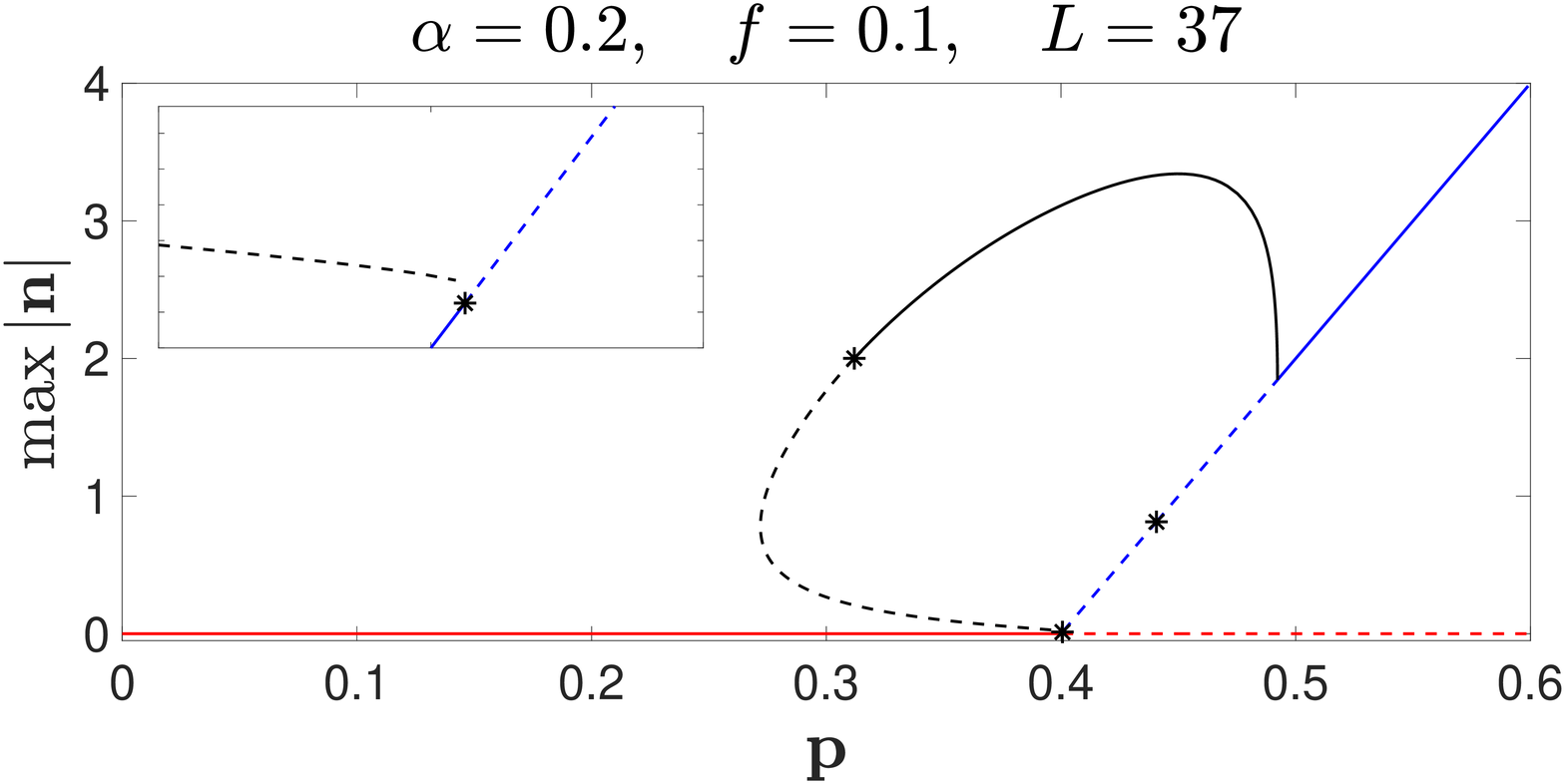}
        \caption*{(b)}
        \qquad
    \end{subfigure}
\qquad
\begin{subfigure}[c]{0.45\textwidth}
        \includegraphics[width=\textwidth, keepaspectratio=true]{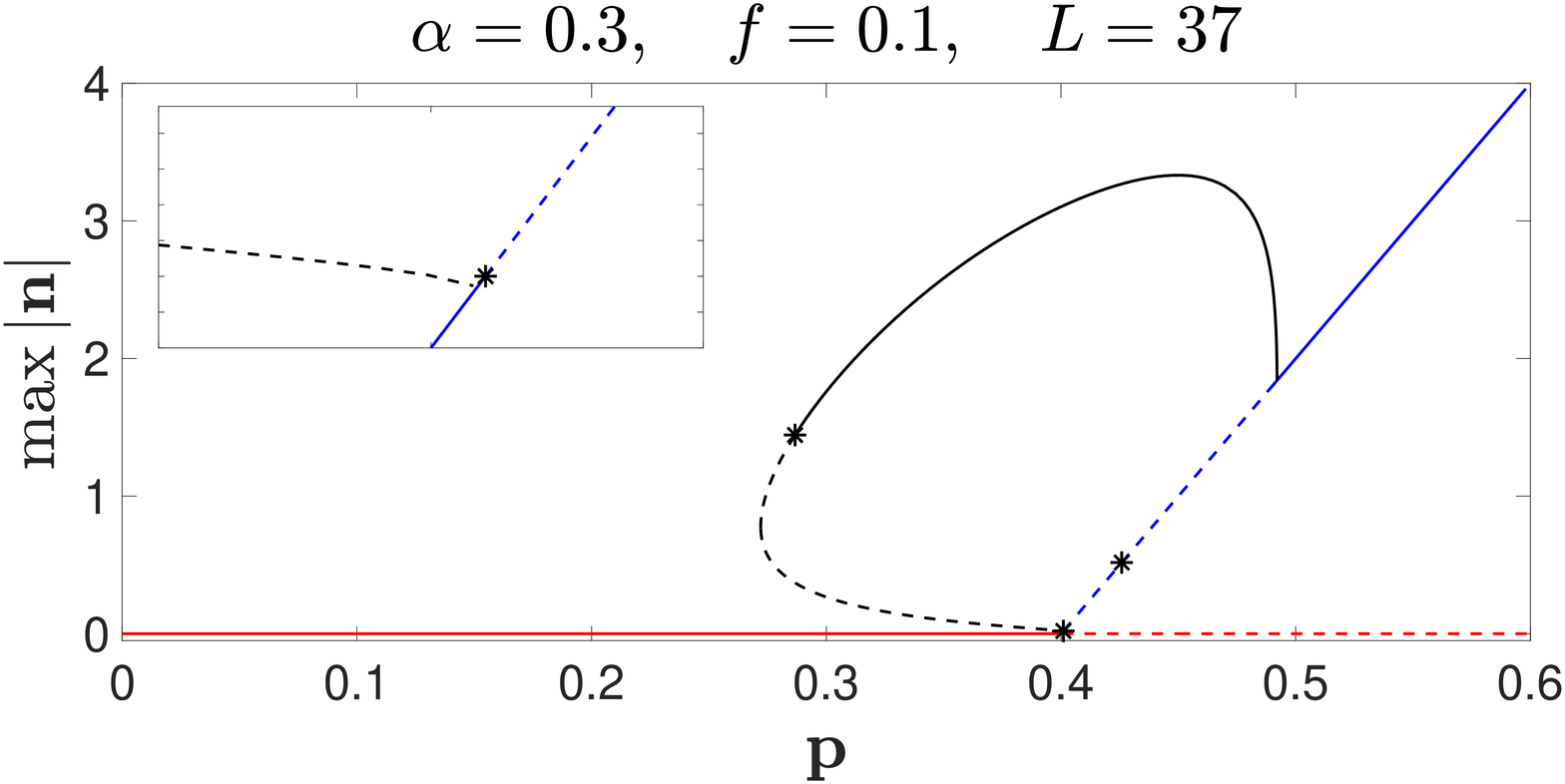}
        \caption*{(c)}
    \end{subfigure}
\qquad
\begin{subfigure}[c]{0.45\textwidth}
        \includegraphics[width=\textwidth, keepaspectratio=true]{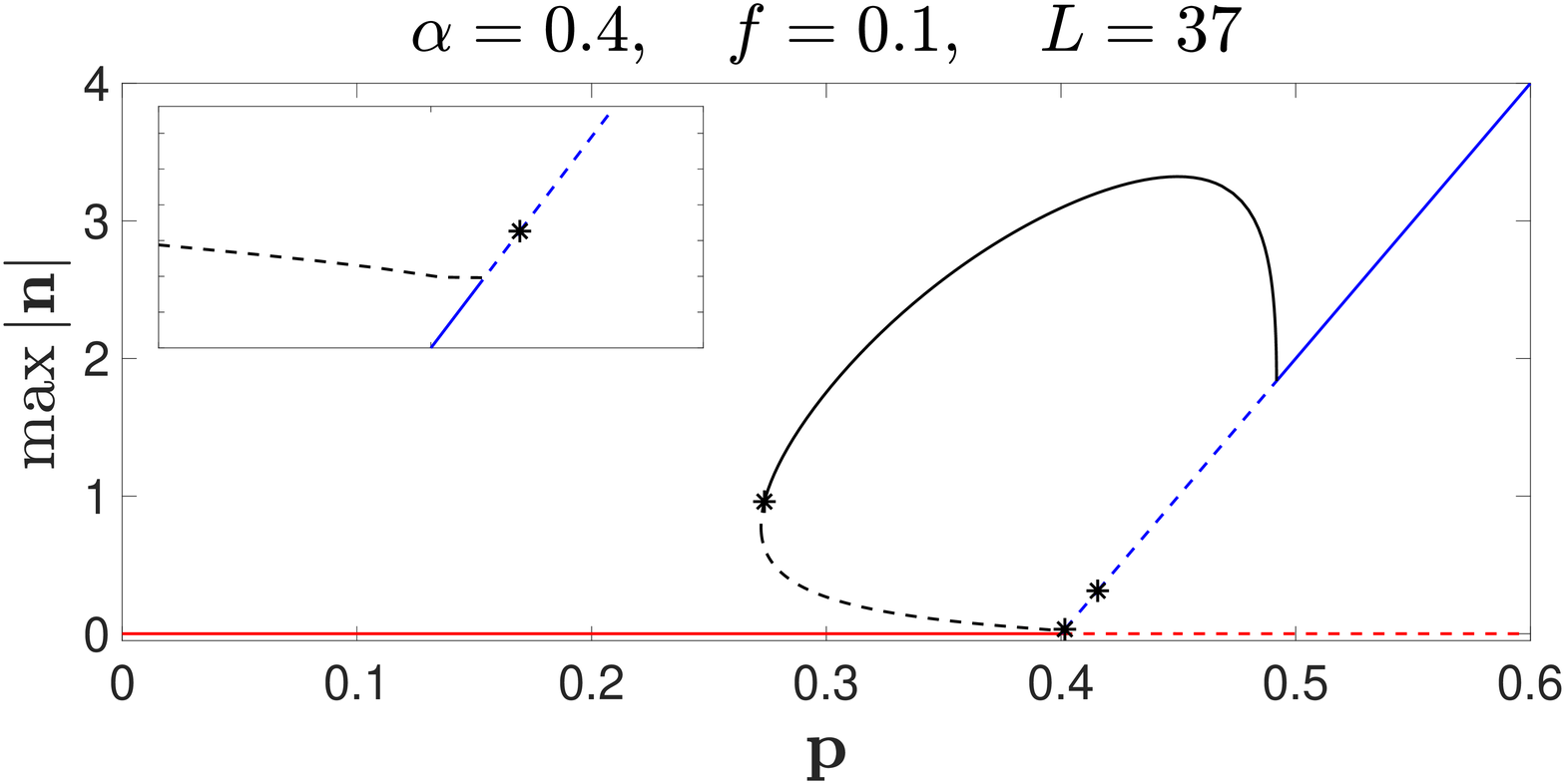}
        \caption*{(d)}
    \end{subfigure}
\caption{Series of bifurcation diagrams, 
  using Fig.~(\ref{fig:Fig1}) as the baseline,
  showing changes caused from small variations in $\alpha$.}\label{fig:Fig2}
\end{figure}

\begin{figure} 
\centering
\begin{subfigure}[c]{0.45\textwidth}
        \includegraphics[width=\textwidth, keepaspectratio=true]{Fig2a.eps}
        \caption*{(a)}
    \end{subfigure}
    \qquad
\begin{subfigure}[c]{0.45\textwidth}
        \includegraphics[width=\textwidth, keepaspectratio=true]{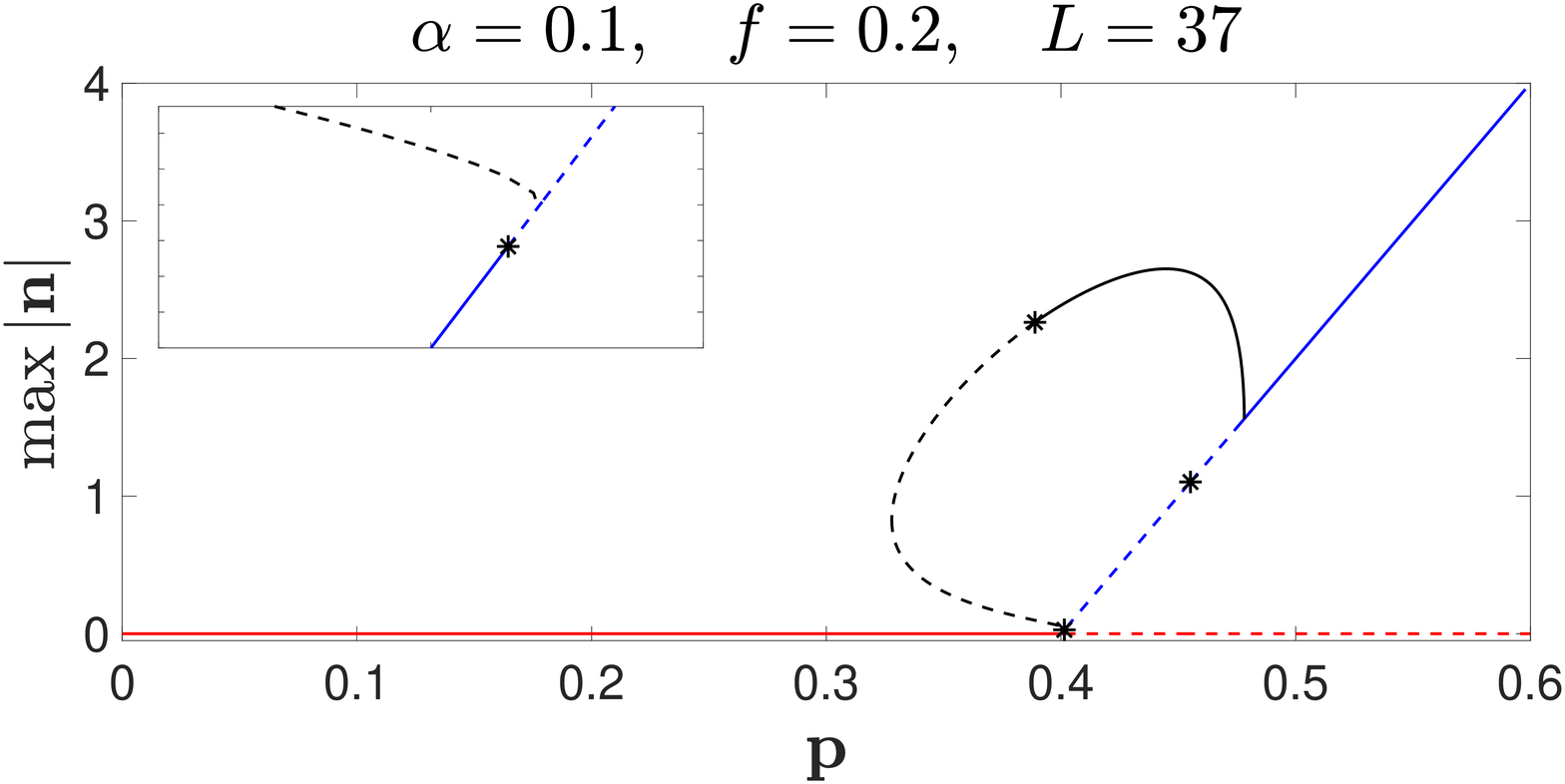}
        \caption*{(b)}
        \qquad
    \end{subfigure}
\qquad
\begin{subfigure}[c]{0.45\textwidth}
        \includegraphics[width=\textwidth, keepaspectratio=true]{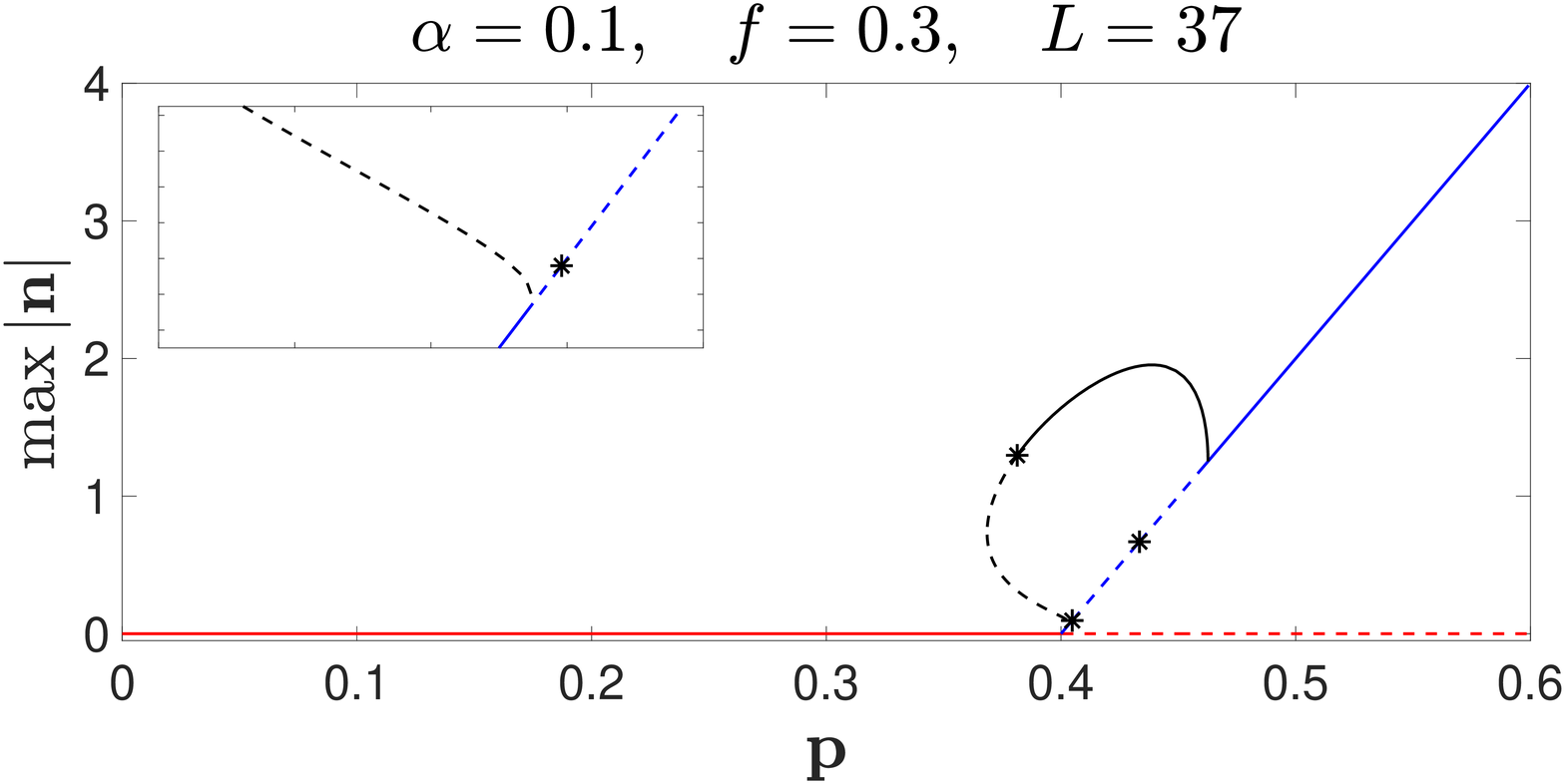}
        \caption*{(c)}
    \end{subfigure}
\qquad
\begin{subfigure}[c]{0.45\textwidth}
        \includegraphics[width=\textwidth, keepaspectratio=true]{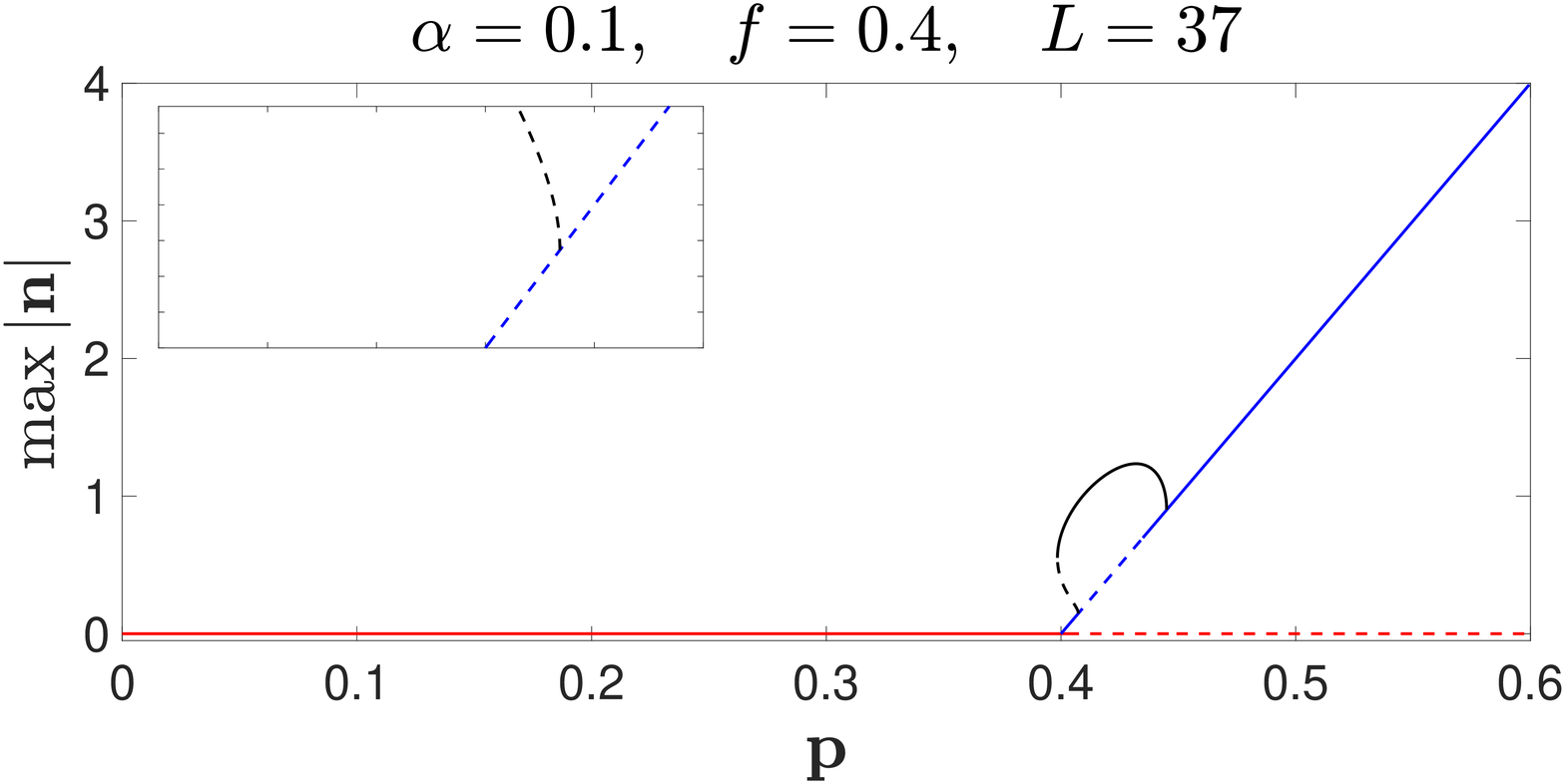}
        \caption*{(d)}
    \end{subfigure}
    \caption{Series of bifurcation diagrams showing changes caused from small variations in $f$.} \label{fig:Fig3} 
\end{figure}

In this section we explore the RM on a domain of size $L=37$. This was chosen,
via Turing analysis, because at most one patterned state bifurcates out of
the homogeneous vegetated state. Fig.~(\ref{fig:Fig2}) and Fig.~(\ref{fig:Fig3}) each contain four bifurcation diagrams corresponding to different values of $\alpha$ and $f$. As a reference point, Fig.~(\ref{fig:Fig1}) is actually the corresponding bifurcation diagram where we have set $\alpha=0.4$ and $f=0.2$, the values used in~\cite{26}. 
We will show that even for this small domain, small
changes in $\alpha$ and $f$ lead to different qualitative behaviors. On larger domains the differences are more striking, as we will see later.
Fig.~(\ref{fig:Fig2}) looks at what changes occur in the bifurcation diagram due to small variations in $\alpha$.
Fig.~(\ref{fig:Fig3}) on the other hand
looks at what changes occur under small variations of $f$.

For concreteness, we first describe the continuation process in detail for Fig.~(\ref{fig:Fig2}a), and then we will discuss the other diagrams more generally. We start on the homogeneous desert branch at $p=0$ and then do continuation in the direction of increasing $p$. At $p=\frac{\mu \nu}{1-\mu}=0.4$, we find the transcritical bifurcation between the homogeneous steady states (the desert
and the vegetated one). Beyond this bifurcation, the desert state actually
has several more branch points. However, we will discuss these in the next
section since they all turn out to be unphysical. Starting at $p=0.4$ on the
homogeneous vegetated branch, we slightly increase $p$ and find a Hopf bifurcation which we have marked with a star. In contrast to Fig.~(\ref{fig:Fig1}) the vegetated branch loses its stability here {\it not} to a subcritical pitchfork. A branch point does occur very soon (parametrically)
after the Hopf bifurcation ($p \approx 0.4001$). This bifurcation is a pitchfork, although both branches are unstable, given that
the ``parent'' branch of the bifurcation was already unstable
and is picking an additional unstable eigendirection.
This can be seen in the inset of the upper left corner of Fig.~(\ref{fig:Fig2}a). Note that we are plotting $p \; \text{vs} \; \max |n|$,
in the relevant bifurcation diagram.
Increasing $p$ further,
we find a second Hopf bifurcation at $p \approx 0.47$. At $p \approx 0.49$
we also identify a supercritical pitchfork. In particular, the vegetated state once again becomes stable and by the same mechanism as that of the upper Turing point in Fig.~(\ref{fig:Fig1}). Upon switching to the patterned
(spatially inhomogeneous solution) branch at $p \approx 0.49$, we find that this branch continues to exist as we
decrease $p$ until we reach a Hopf bifurcation at $p \approx 0.38$. Continuing
to follow this branch, we reach the lower pitchfork.
A similar procedure was carried out for the other bifurcation diagrams and
similar features were identified.

In Fig.~(\ref{fig:Fig2}) we have fixed $f=0.1$ and allowed $\alpha$ to vary from $0.1 \leq \alpha \leq 0.4$, in  increments of 0.1.
In contrast to Fig.~(\ref{fig:Fig1}), we see that Hopf bifurcations are present in both the vegetated (spatially homogeneous) and patterned
(spatially inhomogeneous) states for all of the diagrams. Further, $\alpha$ seems to control the approximate location of the Hopf bifurcations
on the individual branches. As $\alpha$ increases, we see that the two
Hopfs on the vegetated branches seem to approach each other, while on the patterned branch the Hopf seems to approach the fold point. It is interesting to note that, through this mechanism, $\alpha$ also controls the stability of the
branches. For the vegetated branch, we see that the lower branch point
occurs before the Hopf for diagrams Fig.~(\ref{fig:Fig2}c) and Fig.~(\ref{fig:Fig2}d).
However, for Fig.~(\ref{fig:Fig2}a) and Fig.~(\ref{fig:Fig2}b), the vegetated branch loses stability before the branch point, and the pitchfork only renders the solution more unstable.
On the patterned branch, we see a striking contrast between the regions of stability in the diagram: narrow in Fig.~(\ref{fig:Fig2}a) and wide in
Fig.~(\ref{fig:Fig2}d). In fact, taking $\alpha$ less than 0.06, the Hopf bifurcation (on the patterened branch) will move to a $p$-value
greater than 0.4; in other words, when this occurs none of the steady states will be stable in a small interval after $p=0.4$.
We will return to this later, but see Fig.~(\ref{fig:Fig13}) for a
relevant bifurcation diagram. 

In Fig.~(\ref{fig:Fig3}), we have fixed $\alpha=0.1$ and allowed $f$ to vary between $0.1$ and $0.4$, again in increments of $0.1$.
Unlike with variations in $\alpha$, variations in $f$ seem to control the size of the patterned branch.  In particular, while the patterned branch is both wide and tall in Fig.~(\ref{fig:Fig3}a), by Fig.~(\ref{fig:Fig3}d) its deviation from the homogeneous
vegetated branch has been drastically reduced both in height and width
as $f$ increases. Recalling that $n$ is the biomass, this effectively says that that patterned states will have a smaller biomass density at high $f$ values. We see that the lower branch point and the lower Hopf bifurcation point
also switch positions in Fig.~(\ref{fig:Fig3}c) before disappearing
completely in Fig.~(\ref{fig:Fig3}d). 

\begin{figure} 
\centering
\begin{subfigure}[c]{0.45\textwidth}
        \includegraphics[width=\textwidth, keepaspectratio=true]{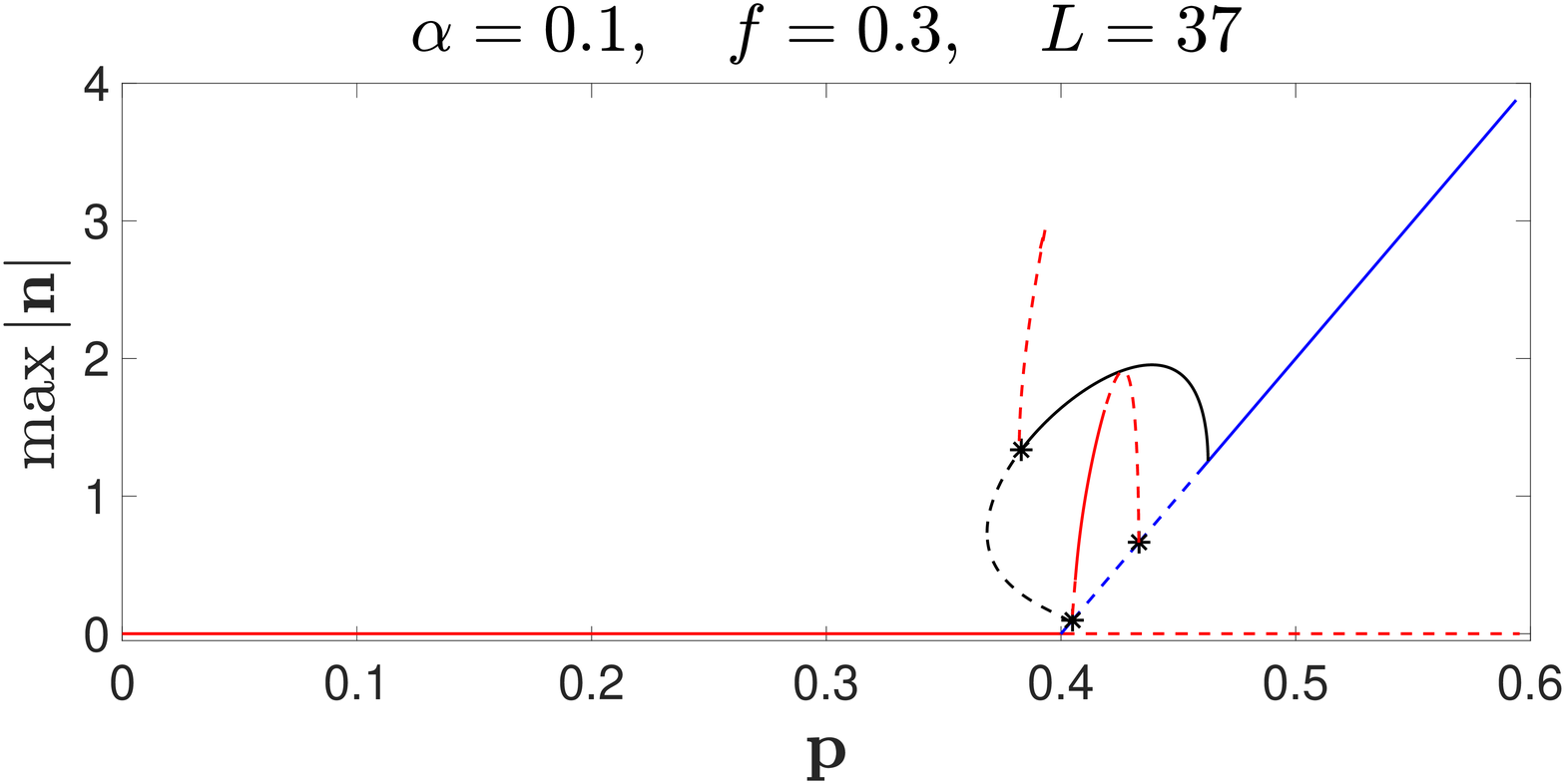}
        \caption*{(a) Bifurcation Diagram}
    \end{subfigure}
    \quad
    \begin{subfigure}[c]{0.45\textwidth}
        \includegraphics[width=\textwidth, keepaspectratio=true]{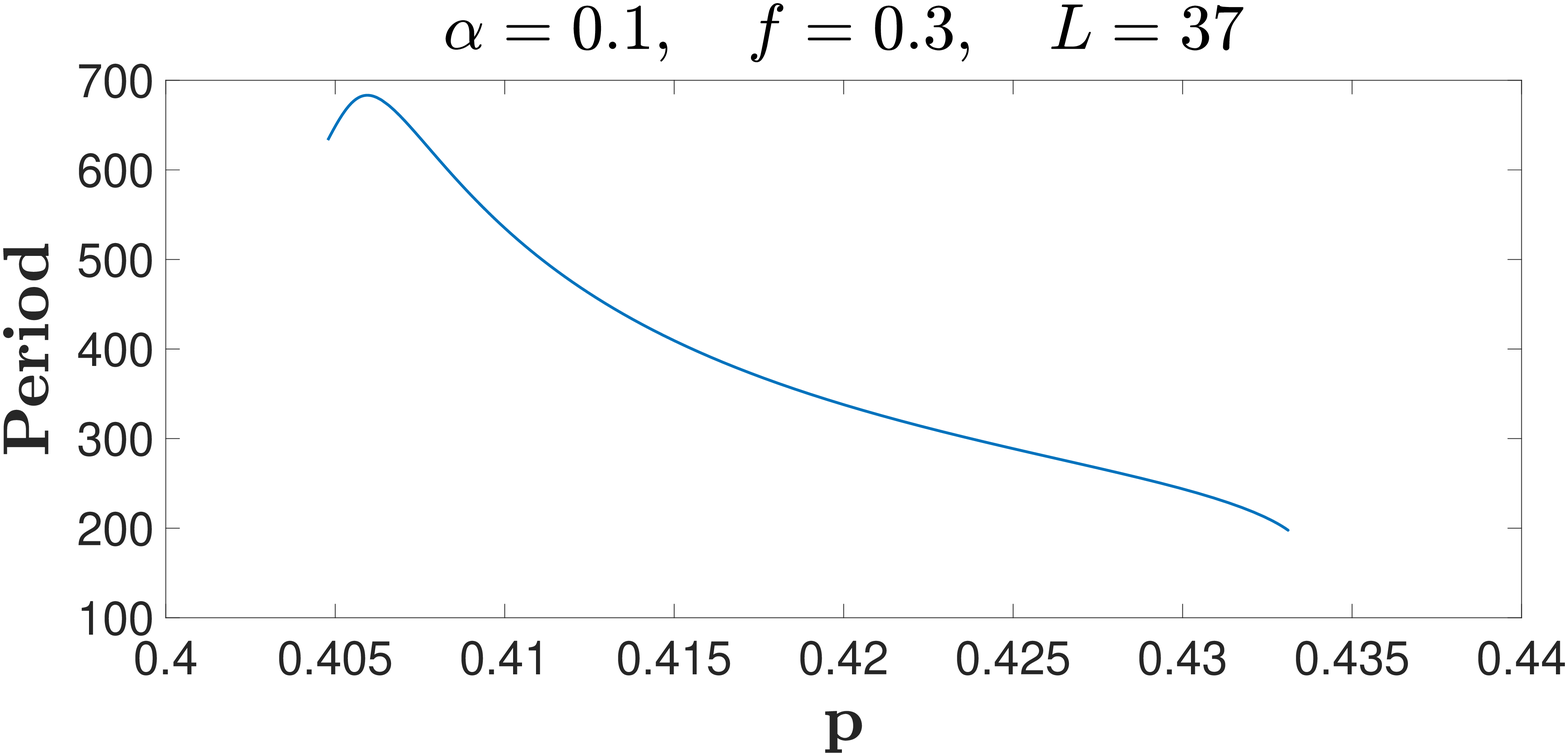}
        \caption*{(b) Period vs. Precipitation}
    \end{subfigure}
    \quad
    \begin{subfigure}[c]{0.45\textwidth}
        \includegraphics[width=\textwidth, keepaspectratio=true]{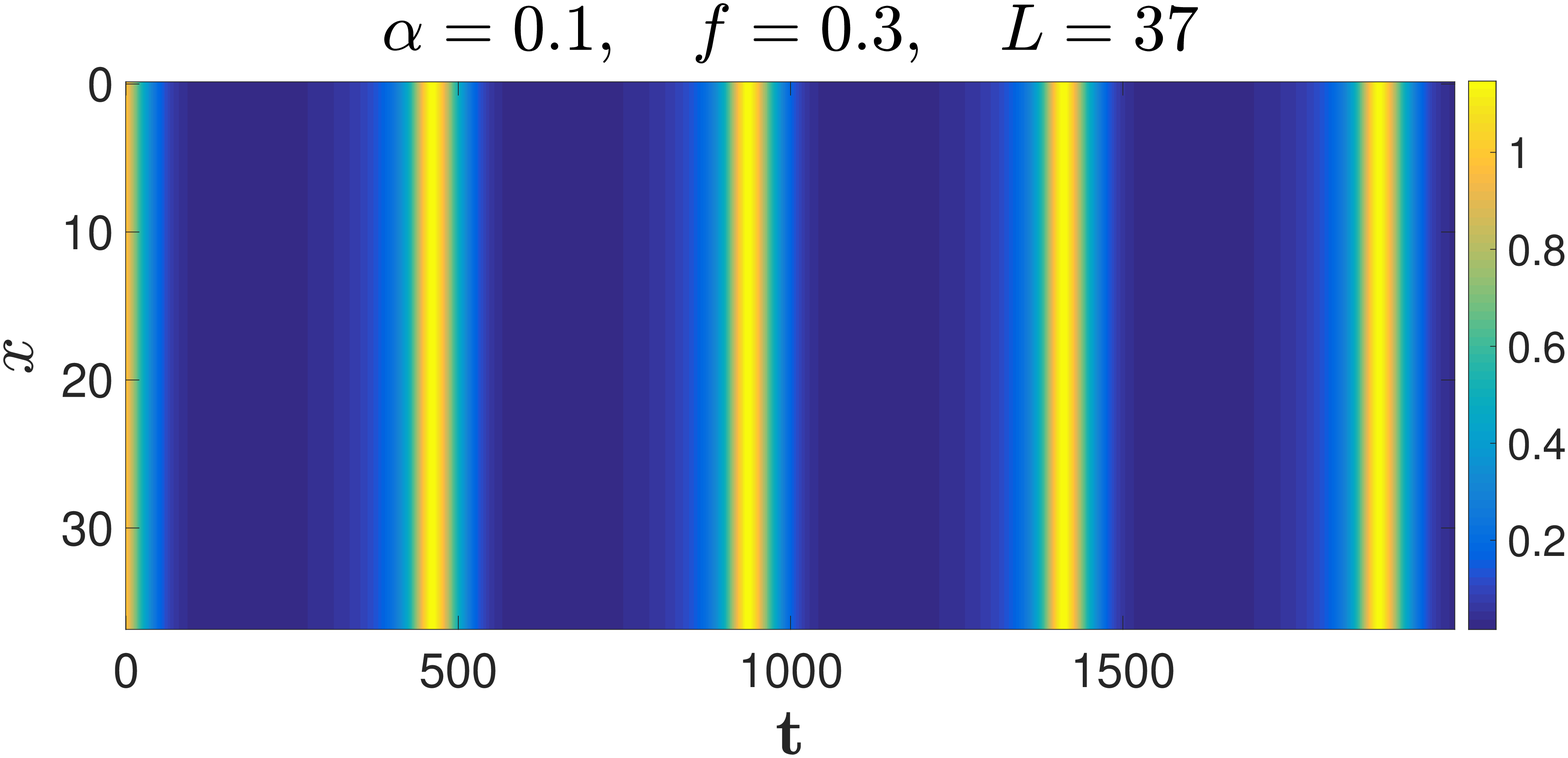}
        \caption*{(c) Biomass}
    \end{subfigure}
    \quad
    \caption{(a) Bifurcation diagram including the
      periodic orbits. (b) Period vs precipitation for the periodic orbit emanating from the vegetated branch. (c) Example of the previously mentioned orbit, showing that it is spatially homogeneous.}\label{fig:Fig4}
\end{figure}

  \begin{figure} 
\centering
    \begin{subfigure}[c]{0.45\textwidth}
        \includegraphics[width=\textwidth, keepaspectratio=true]{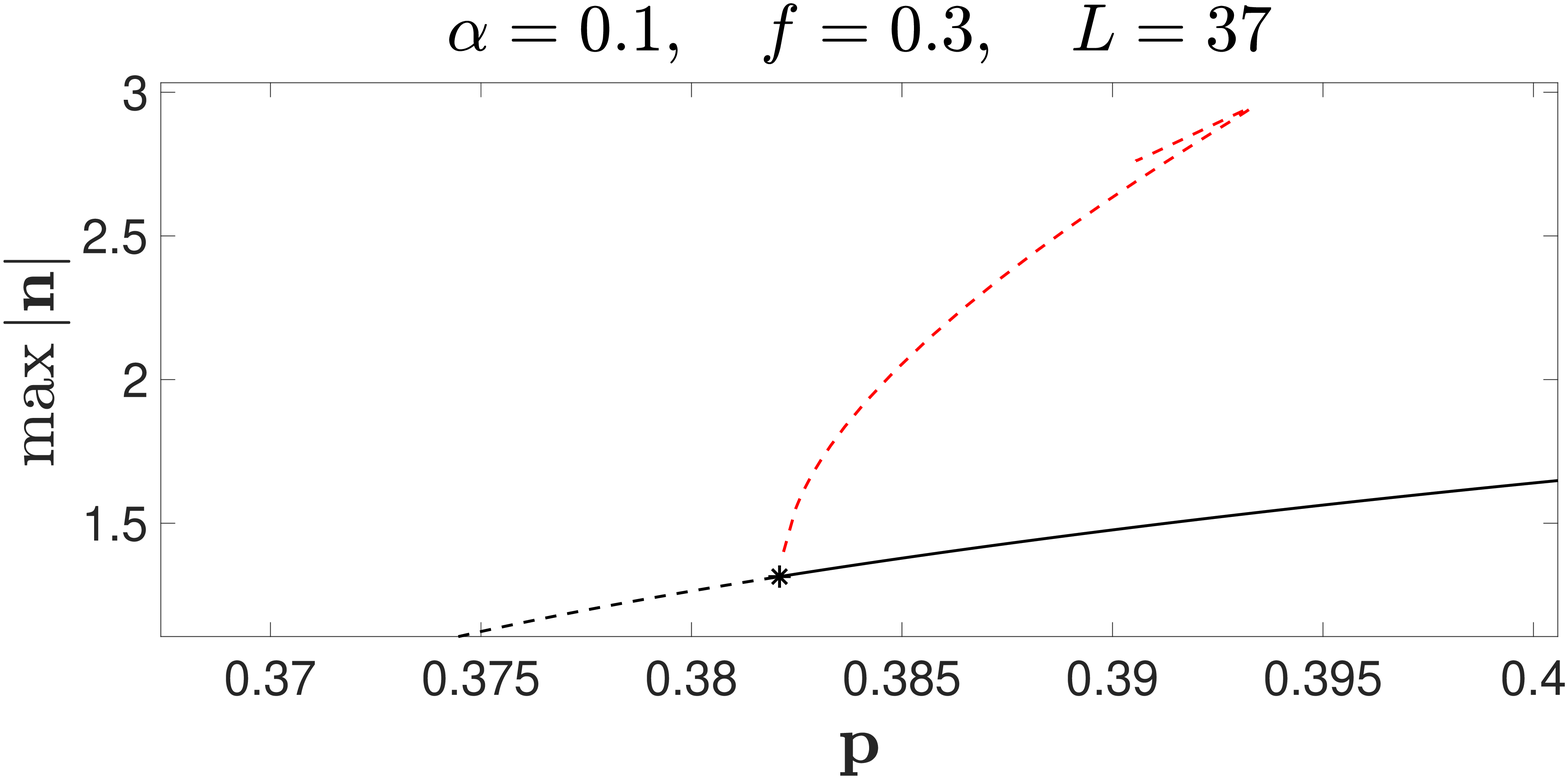}
        \caption*{(a) Bifurcation Diagram}
    \end{subfigure}
    \quad
    \begin{subfigure}[c]{0.45\textwidth}
        \includegraphics[width=\textwidth, keepaspectratio=true]{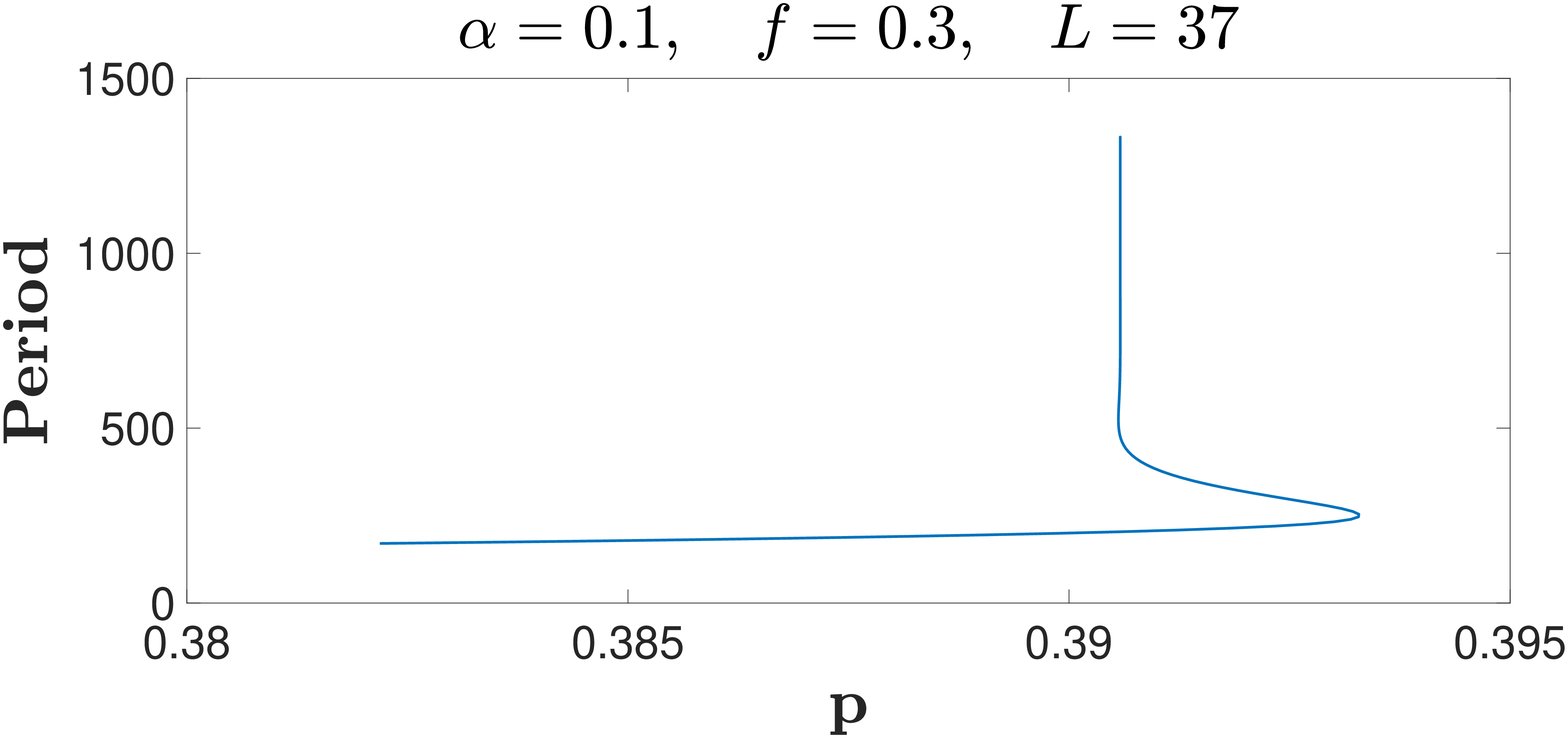}
        \caption*{(b) Period vs. Precipitation}
    \end{subfigure}
    \quad
    \begin{subfigure}[c]{0.45\textwidth}
        \includegraphics[width=\textwidth, keepaspectratio=true]{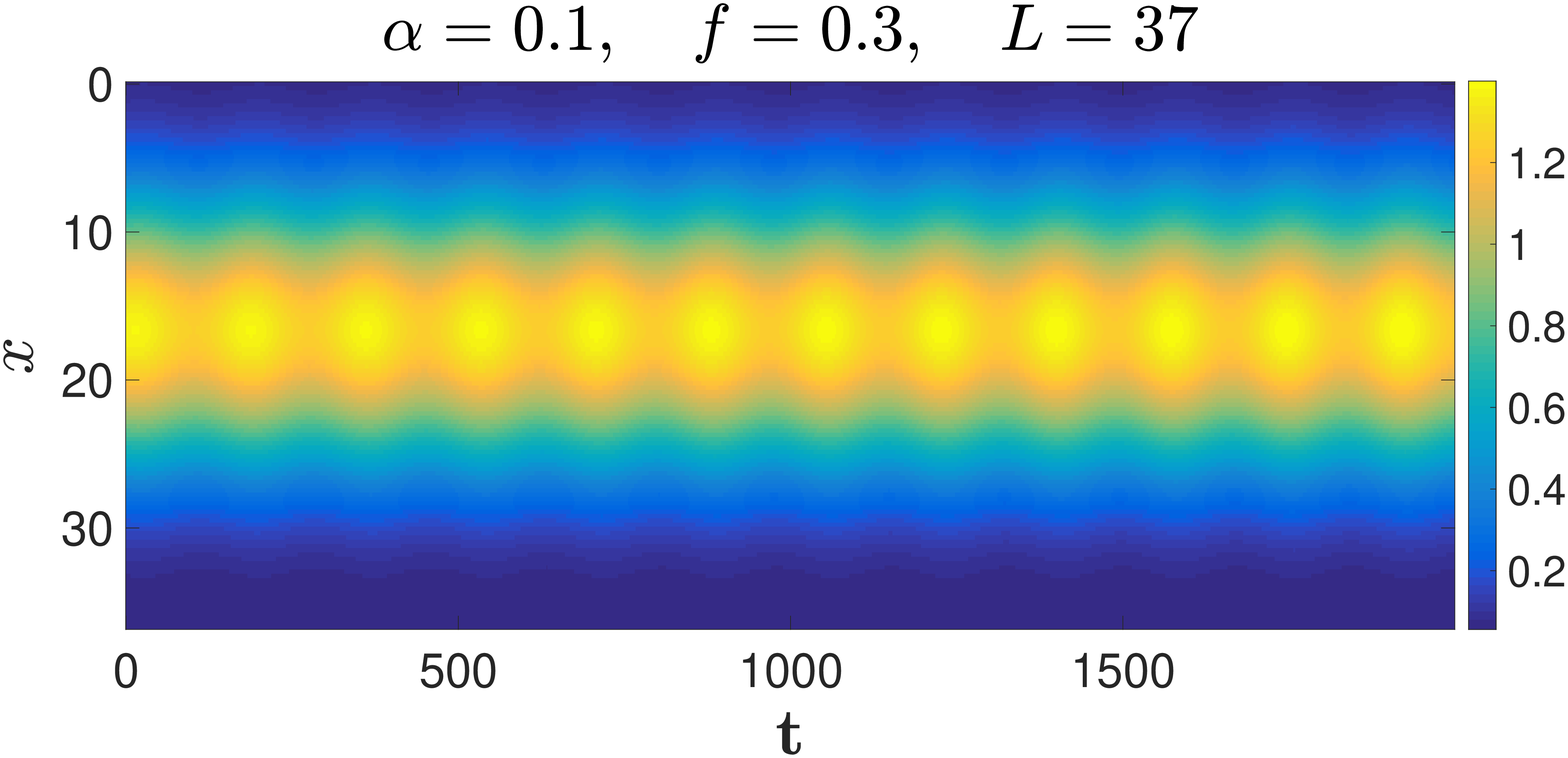}
        \caption*{(c) Biomass}
    \end{subfigure}
    \quad
    \caption{(a) Diagram (\ref{fig:Fig4}a) zoomed in on the periodic orbit emanating from the patterned branch. (b) Period vs precipitation of the aforementioned orbit. Notice that the period seems to increase without bound, indicating a possible homoclinic bifurcation. (c) Contour plot of the previously mentioned orbit at a selected point.}\label{fig:Fig52}
\end{figure}

  \begin{figure} 
\centering
    \begin{subfigure}[c]{0.45\textwidth}
        \includegraphics[width=\textwidth, keepaspectratio=true]{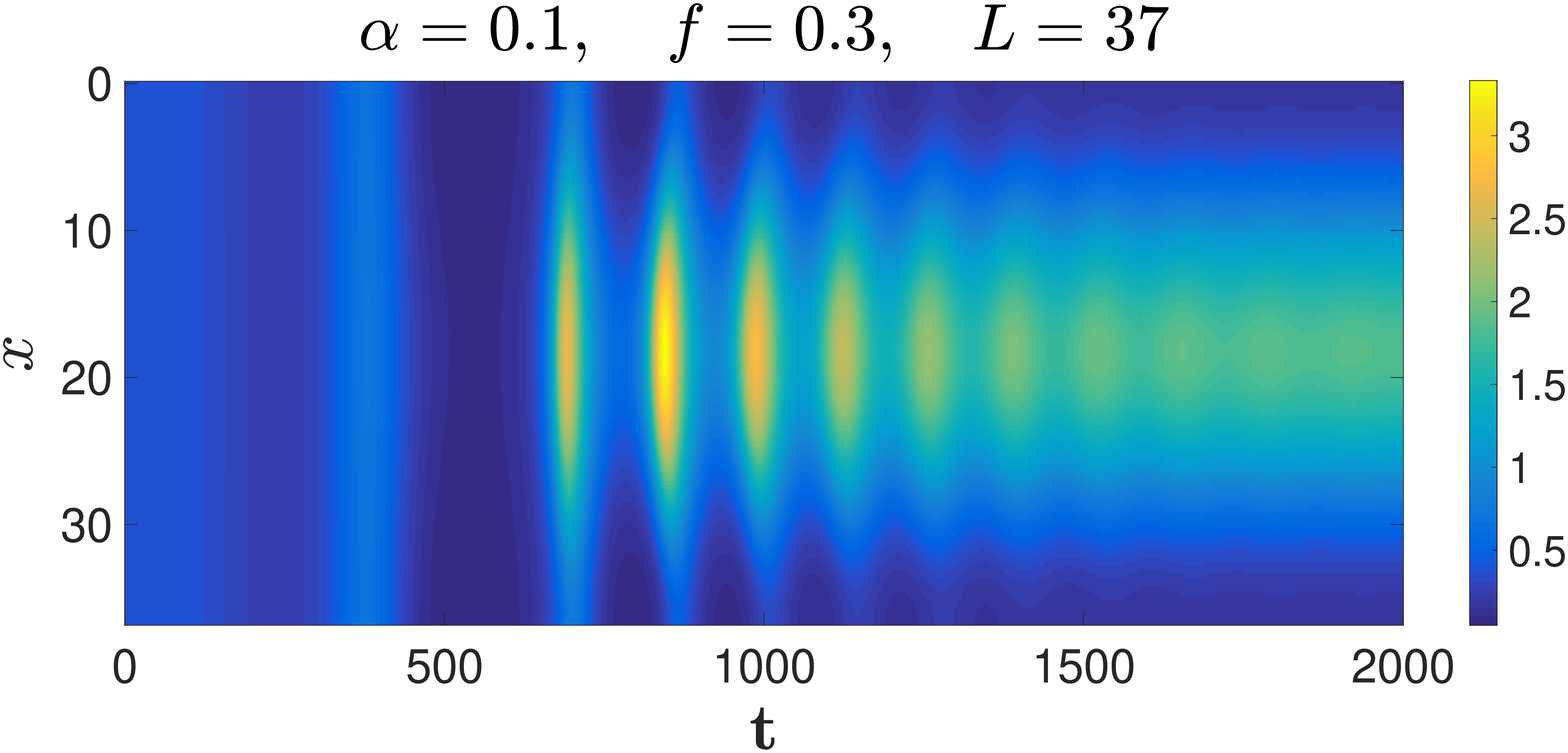}
        \caption*{(a) Generic Perturbation}
    \end{subfigure}
    \quad
    \begin{subfigure}[c]{0.45\textwidth}
        \includegraphics[width=\textwidth, keepaspectratio=true]{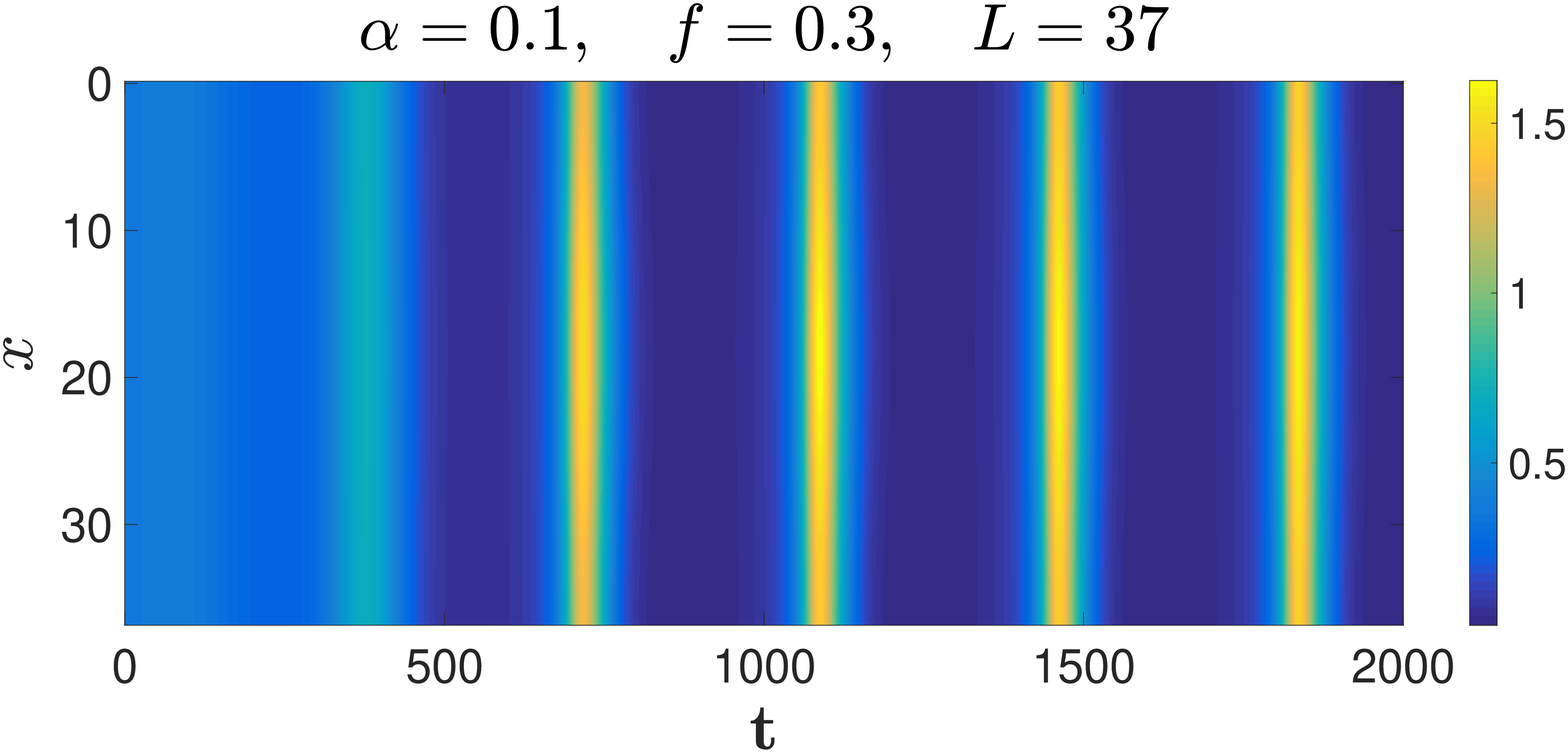}
        \caption*{(b) Nongeneric Perturbation}
    \end{subfigure}
    \caption{Contour plots showing the evolution of the system from a perturbed
    point on the vegetated branch (a) to a point on the patterned branch or (b) to the homogeneous periodic orbit. We have fixed $p$ at 0.417.}\label{fig:Fig42}
\end{figure}

Using MATCONT \cite{7,8,9,10,11}, we continue the periodic orbits emanating from the Hopf bifurcations in Fig.~(\ref{fig:Fig3}c); Fig.~(\ref{fig:Fig4}a) shows the corresponding bifurcation diagram. Because of the large number of equations (due to the spatial discretization), we were unable to obtain the multipliers directly from MATCONT. Instead, we perturbed each point along the periodic orbit with a inhomogeneous perturbation (containing the full spectrum of fourier modes on the given grid) and then integrated this initial condition for $10^5$ time units; if the resulting numerical solution converged back to the periodic orbit on this time span, then we marked it as stable in Fig.~(\ref{fig:Fig4}a). We mention that the integration was done via the ETDRK4 integrator~\cite{14,15}, as the system is inherently stiff.

In Fig.~(\ref{fig:Fig4}a), notice that the periodic orbit connects the two Hopf bifurcation points  on the homogeneous vegetated branch  and is stable for a large
range of $p$-values.. Furthermore, Fig.~(\ref{fig:Fig4}c) reveals that this periodic orbit is spatially homogeneous. Indeed, below we show that this periodic orbit is present in the system with diffusion removed. 

Fig.~(\ref{fig:Fig52}a) shows the bifurcation diagram Fig.~(\ref{fig:Fig4}a) except zoomed in on the periodic orbit emanating from the patterned state. Contrary to the previous periodic orbit, this cycle is everywhere unstable and inhomogeneous. Fig.~(\ref{fig:Fig52}c) shows a contour plot of this inhomogeneous periodic orbit. In addition,  Fig.~(\ref{fig:Fig52}b) shows that the period appears to increase with out bound as one continues the periodic orbit. This a typical indicator of the existence of a homoclinic orbit and, indeed, it appears that the periodic orbit undergoes a homoclinic bifurcation. Inspecting one of these large period orbits, it appears that it is homoclinic to the unstable portion of the inhomogeneous steady state at the corresponding $p$-value (see Fig.~(\ref{fig:Fig4}a)).

We also studied the dynamics of the system using the ETDRK4 integrator. Our general procedure was to pick a point on the homogeneous
vegetated branch, perturb the point with a homogeneous or inhomogeneous
perturbation, and then use this as our initial condition in the integrator. As might be expected, we found that homogeneous perturbations would always converge to the periodic orbit (for $p$-values between the Hopf bifurcations). Likewise, generic inhomogenous perturbations almost always converge to the patterned state. The exception is when an inhomogeneous pertubation contains a very small inhomogenous part compared to the homogeneous part i.e. the nontrivial Fourier modes have extremely small amplitudes compared to the constant factor. In this latter case, the system would indeed evolve to the homogeneous periodic orbit. Examples can be found in Fig.~(\ref{fig:Fig42}). 

With an eye towards the underlying ecology, we remark that if the system is in the desert or patterned state and a rapid change in the precipitation parameter puts it into the region between the Hopf bifurcations (on the vegetated branch), then indeed the system will evolve to the homogenous periodic orbit. Thus, there is a (generic) multistability and \emph{potential}  hysteresis within the system between the nonhomogenous steady state and the homogeneous periodic orbit.

\begin{figure}
\centering
\begin{subfigure}[c]{0.5\textwidth}
        \includegraphics[width=\textwidth, keepaspectratio=true]{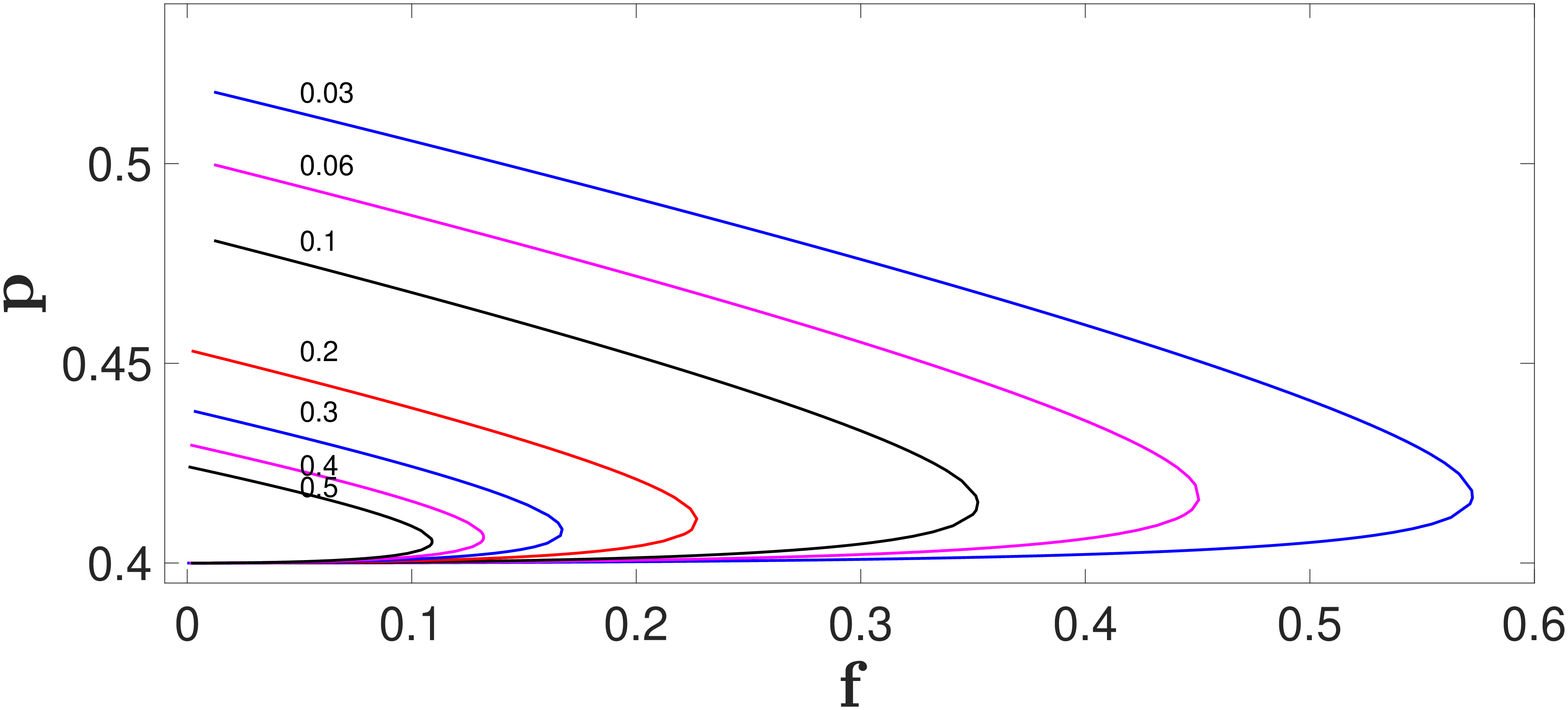}
        \caption*{(a) Two parameter continuation of the Hopf bifurcations on the vegetated branch in the plane of precipitation $p$ vs. infiltration
          parameter $f$.}
    \end{subfigure}
    \quad
    \begin{subfigure}[c]{0.5\textwidth}
        \includegraphics[width=\textwidth, keepaspectratio=true]{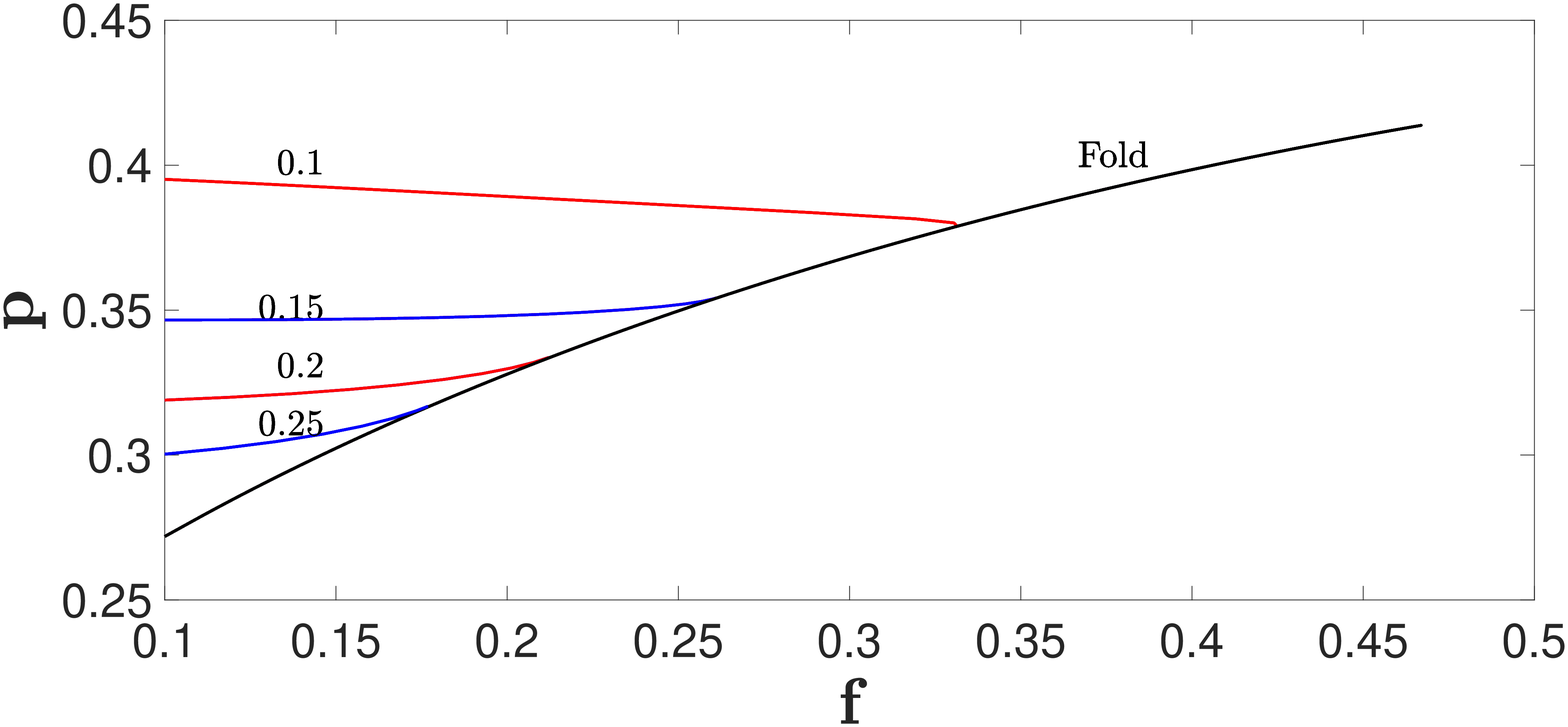}
        \caption*{(b) Two parameter continuation of the Hopf bifurcation on the patterned branch (using the same two parameter plane as above).}
    \end{subfigure}
     \caption{}
     \label{fig:Fig5}
\end{figure}

Returning to the bifurcation diagrams in Fig.~(\ref{fig:Fig2}), we
can perform a two parameter continuation and track the Hopf bifurcations appearing in the diagrams. To do this, we fix $\alpha$ at a given value, then do a continuation, where $p$ and $f$ are allowed to vary, generating a curve in the
$(p,f)$-plane.
Doing this for various values of $\alpha$ generates many such curves.
Fig.~(\ref{fig:Fig5}) shows the locus of these Hopf bifurcations.
Fig.~(\ref{fig:Fig5}a) shows the diagrams for the vegetated branch and Fig.~(\ref{fig:Fig5}b) the patterned branch. The small numbers appearing on the left hand sides are the fixed values of $\alpha$ used in the continuation. In Fig.~(\ref{fig:Fig5}b) the large curve labeled ``Fold" is in fact the locus of the fold bifurcation appearing in the patterned branch. Note that there are actually four such curves plotted (for the respective folds),
each corresponding to a different $\alpha$, but they
are so close together that they are indistinguishable in the plot.

As the diagrams in Fig.~(\ref{fig:Fig2}) and Fig.~(\ref{fig:Fig3}) suggest, the Hopf bifurcations on the vegetated branch do eventually annihilate one another as either $\alpha$ or $f$ increases. Similarly, the Hopf bifurcation on the patterned branch converges to the fold point as either $\alpha$ or $f$ increases. We also note that Hopf bifurcations do not exist in those previous studies which fixed $\alpha=0.4$ and $f=0.2$, as one would expect.

Lastly, we verify in the appendix, via a Turing analysis, that when
the PDE becomes unstable to these Hopf bifurcations on the vegetated branch,
the unstable mode corresponds
to the zero wavenumber (i.e., homogeneous in space mode),
and that this phenomenon is already present in the ODE system
in the absence of diffusion.
This implies that the existence and location of the Hopf
bifurcations on the vegetated branch are independent of the domain size
used and the value of the diffusion coefficients. In particular, they
will be present in the corresponding 2D system, a feature worthwhile
of further consideration in its own right.

\subsection{Increased Domain Size}

\begin{figure}  
\centering
    \begin{subfigure}[c]{0.45\textwidth}
        \includegraphics[width=\textwidth, keepaspectratio=true]{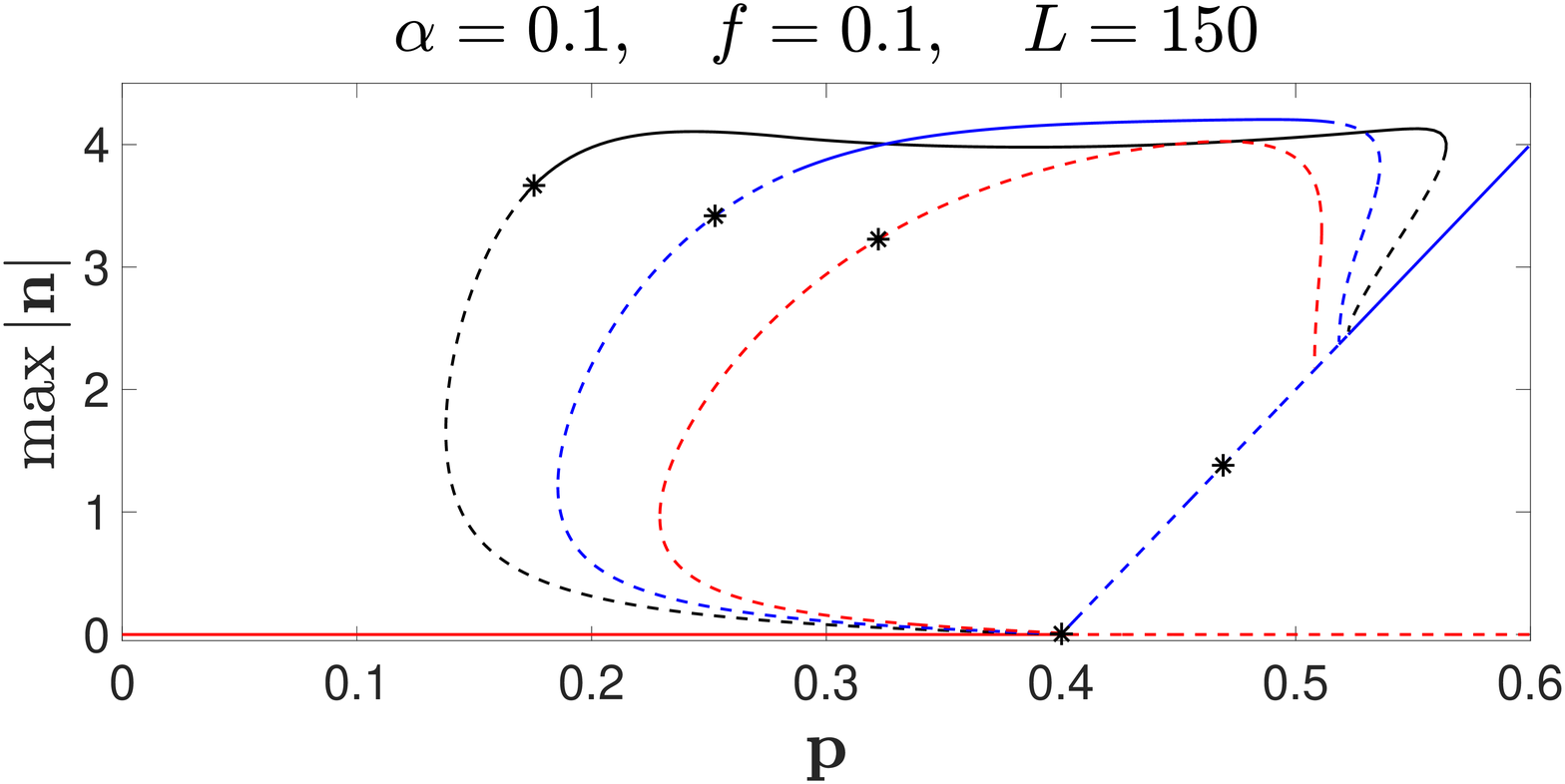}
        \caption*{(a)}
    \end{subfigure}
    \qquad
    \begin{subfigure}[c]{0.45\textwidth}
        \includegraphics[width=\textwidth, keepaspectratio=true]{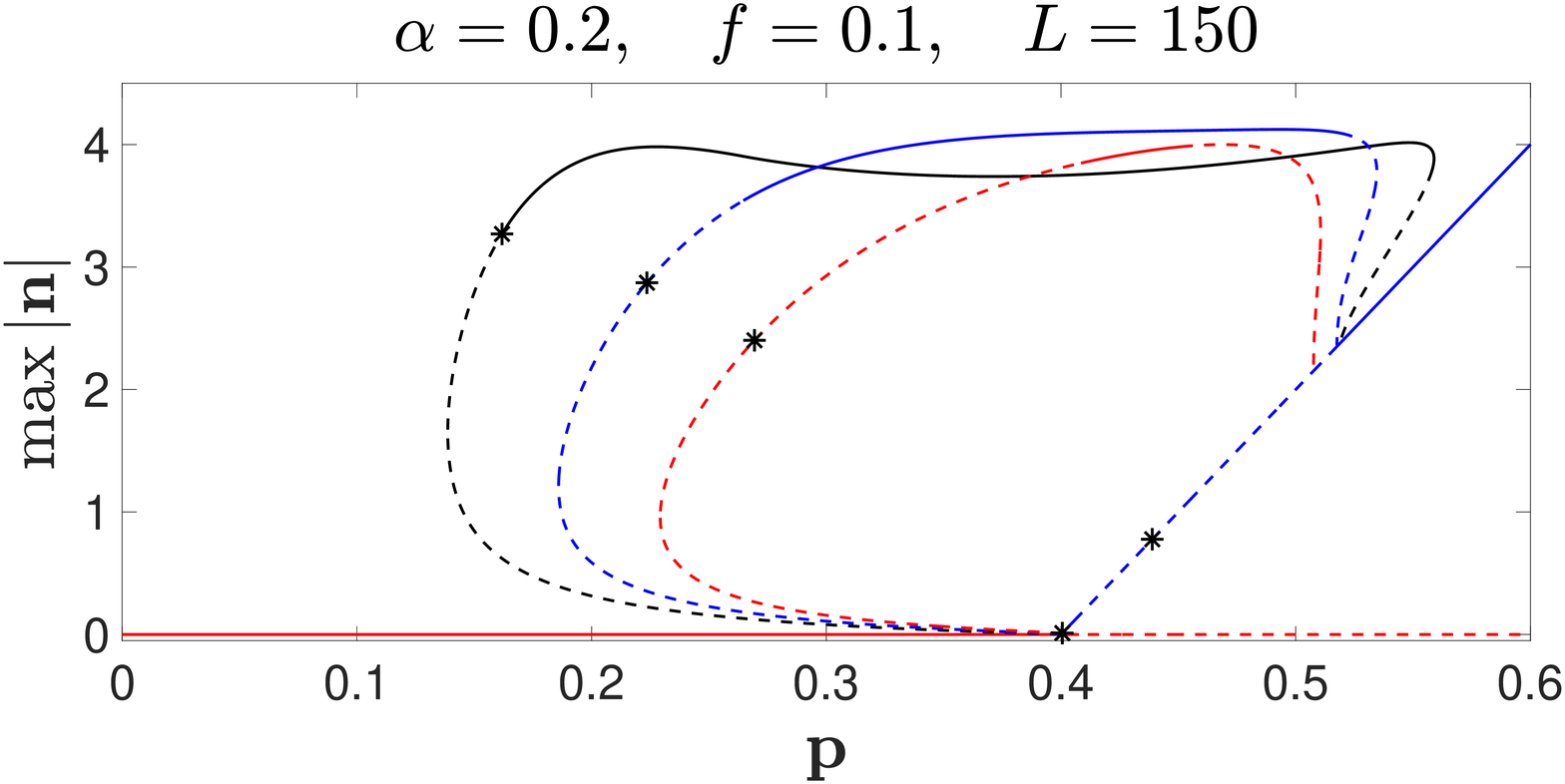}
        \caption*{(b)}
    \end{subfigure}
        \qquad
    \begin{subfigure}[c]{0.45\textwidth}
        \includegraphics[width=\textwidth, keepaspectratio=true]{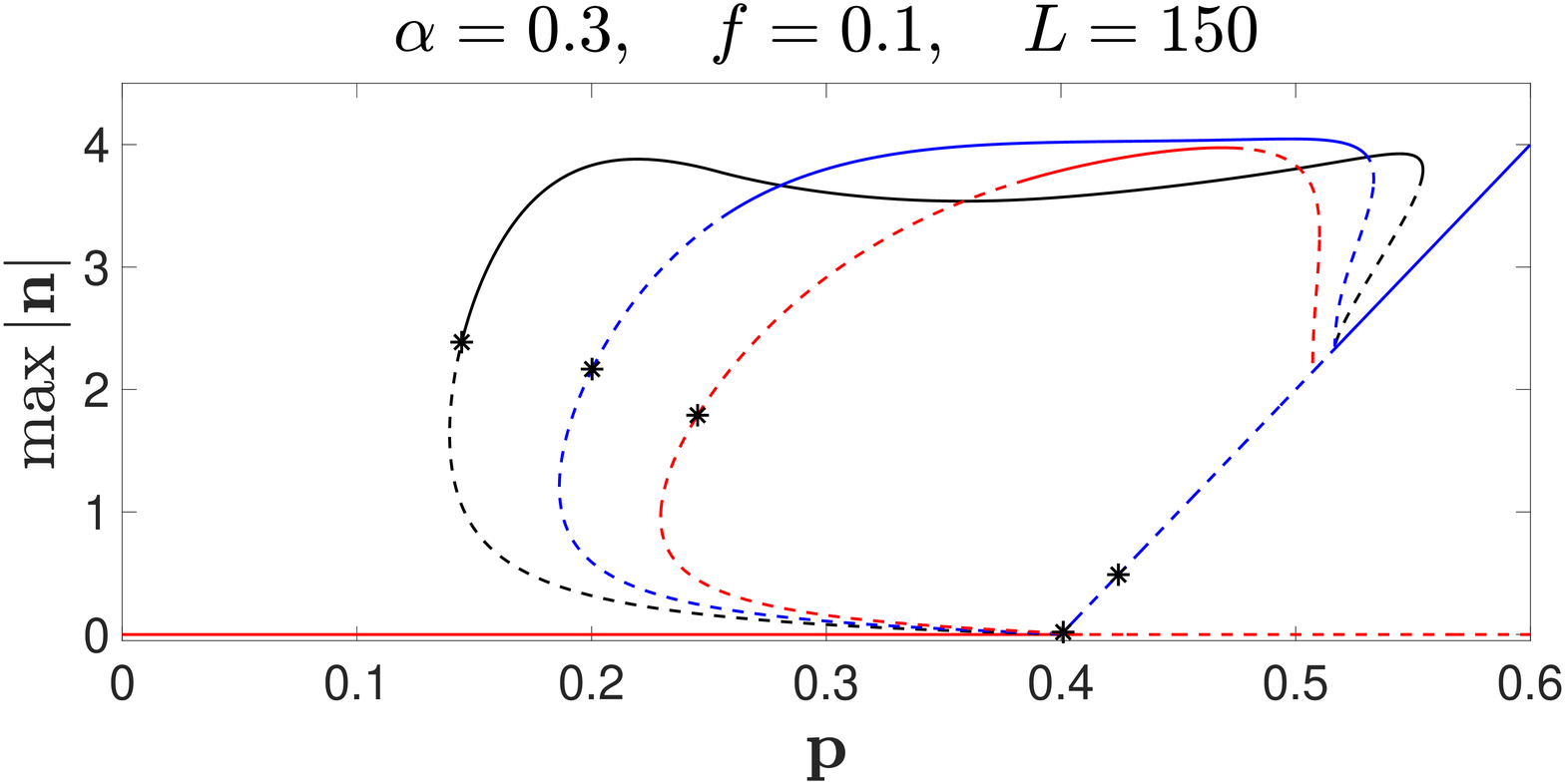}
        \caption*{(c)}
    \end{subfigure}
        \qquad
    \begin{subfigure}[c]{0.45\textwidth}
        \includegraphics[width=\textwidth, keepaspectratio=true]{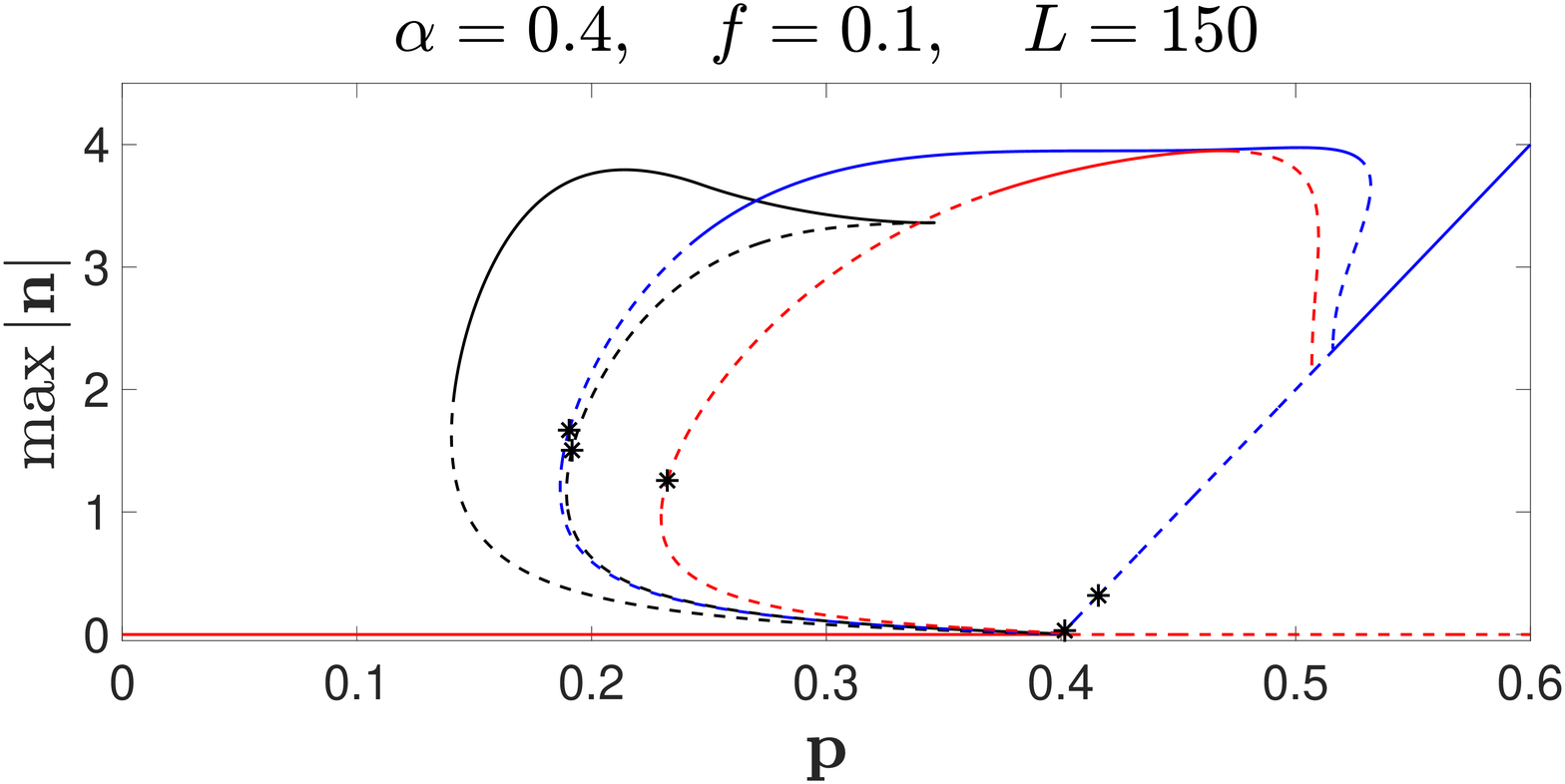}
        \caption*{(d)}
    \end{subfigure}
    \quad
     \begin{subfigure}[c]{0.45\textwidth}
        \includegraphics[width=\textwidth, keepaspectratio=true]{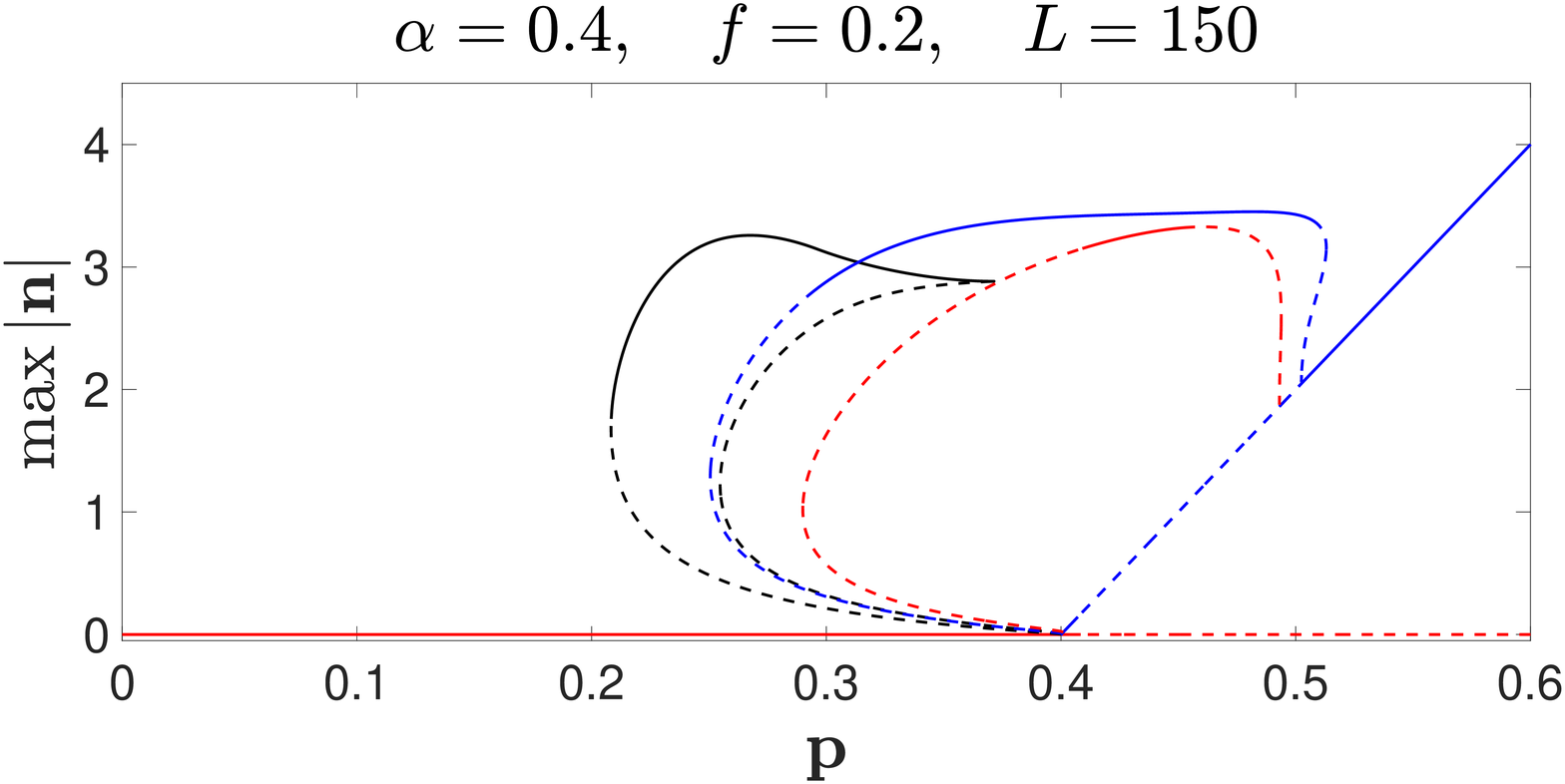}
        \caption*{(e)}
    \end{subfigure}
     \caption{Bifurcation diagrams of the RM, similarly
       to the previous ones, but now for domain length
       $L=150$.} \label{fig:Fig6}
\end{figure}

In the previous section the domain size was chosen so that the bifurcation diagrams would be as simple as possible. In this section, we increase the domain size to $L=150$ so that multiple patterned branches are present. We then construct several bifurcation diagrams allowing the infiltration
parameters $\alpha$ and $f$ to vary. As we will see, Hopf bifurcations do persist on the patterned branches and $\alpha, f$ play an even more striking role in the form that the patterned branches take.

\begin{figure}  
\centering
\begin{subfigure}[c]{0.45\textwidth}
        \includegraphics[width=\textwidth, keepaspectratio=true]{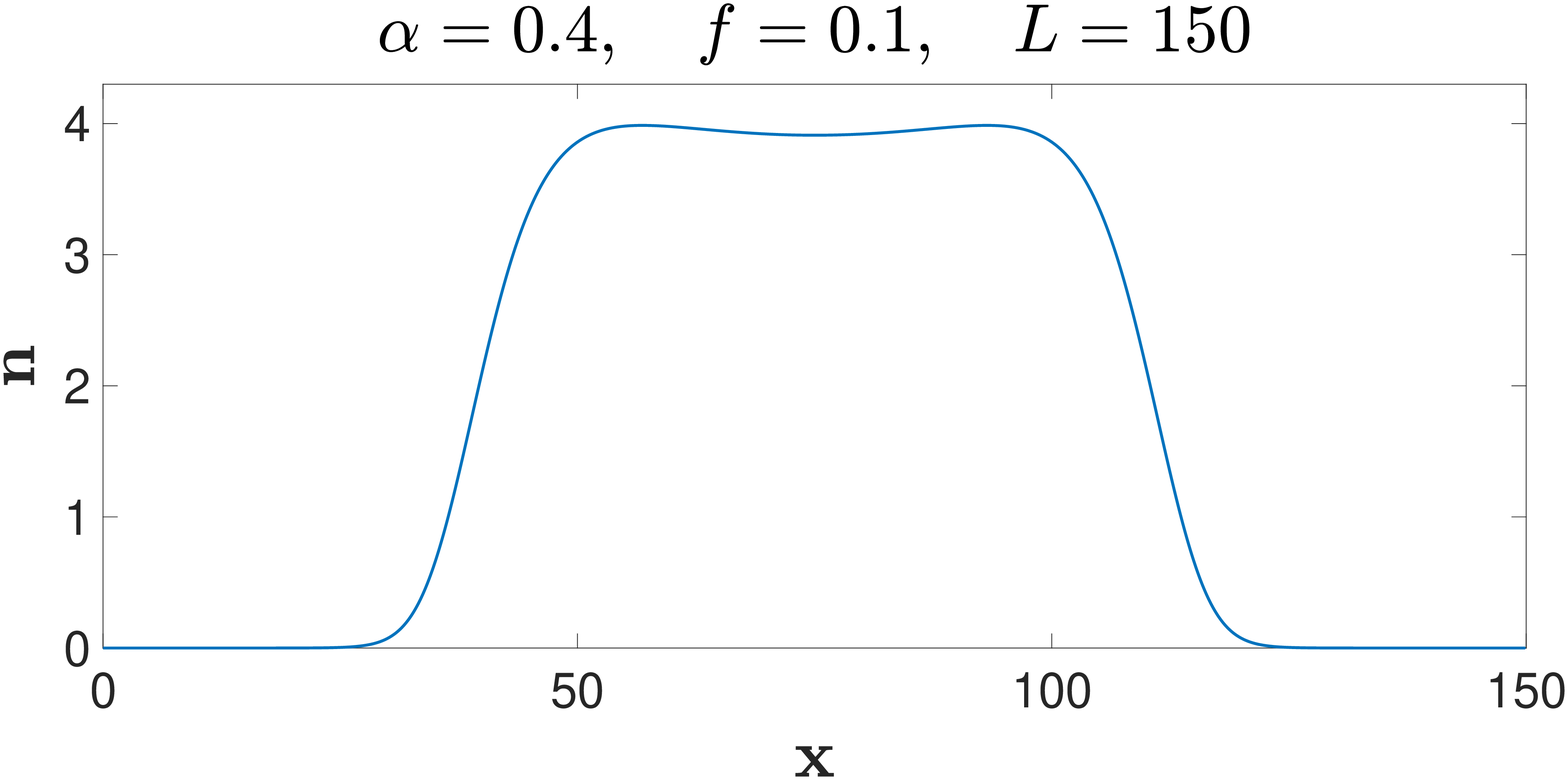}
        \caption*{(a) Black Branch}
    \end{subfigure}
    \quad
    \begin{subfigure}[c]{0.45\textwidth}
        \includegraphics[width=\textwidth, keepaspectratio=true]{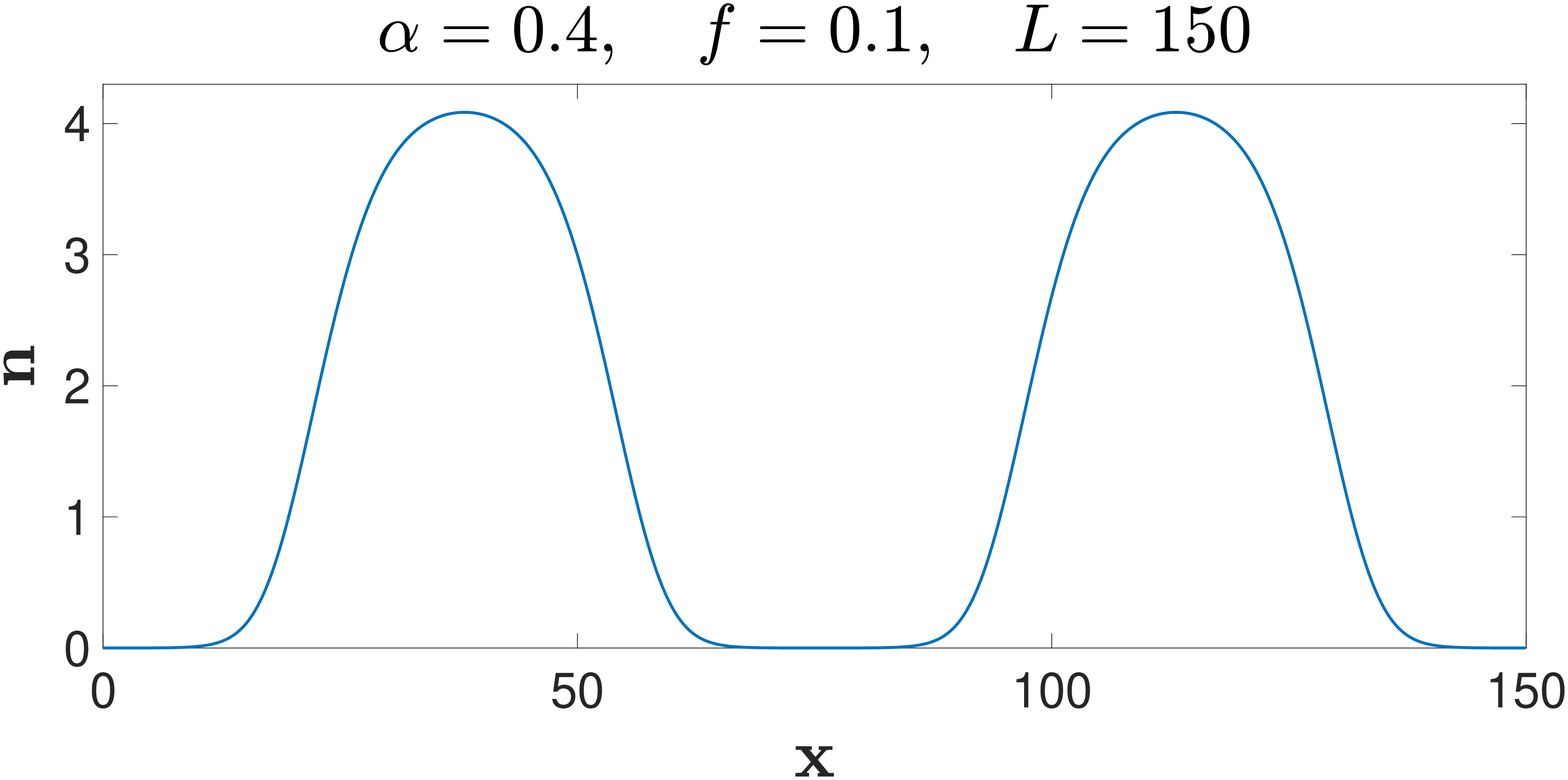}
        \caption*{(b) Blue Branch}
    \end{subfigure}
    \quad
    \begin{subfigure}[c]{0.45\textwidth}
        \includegraphics[width=\textwidth, keepaspectratio=true]{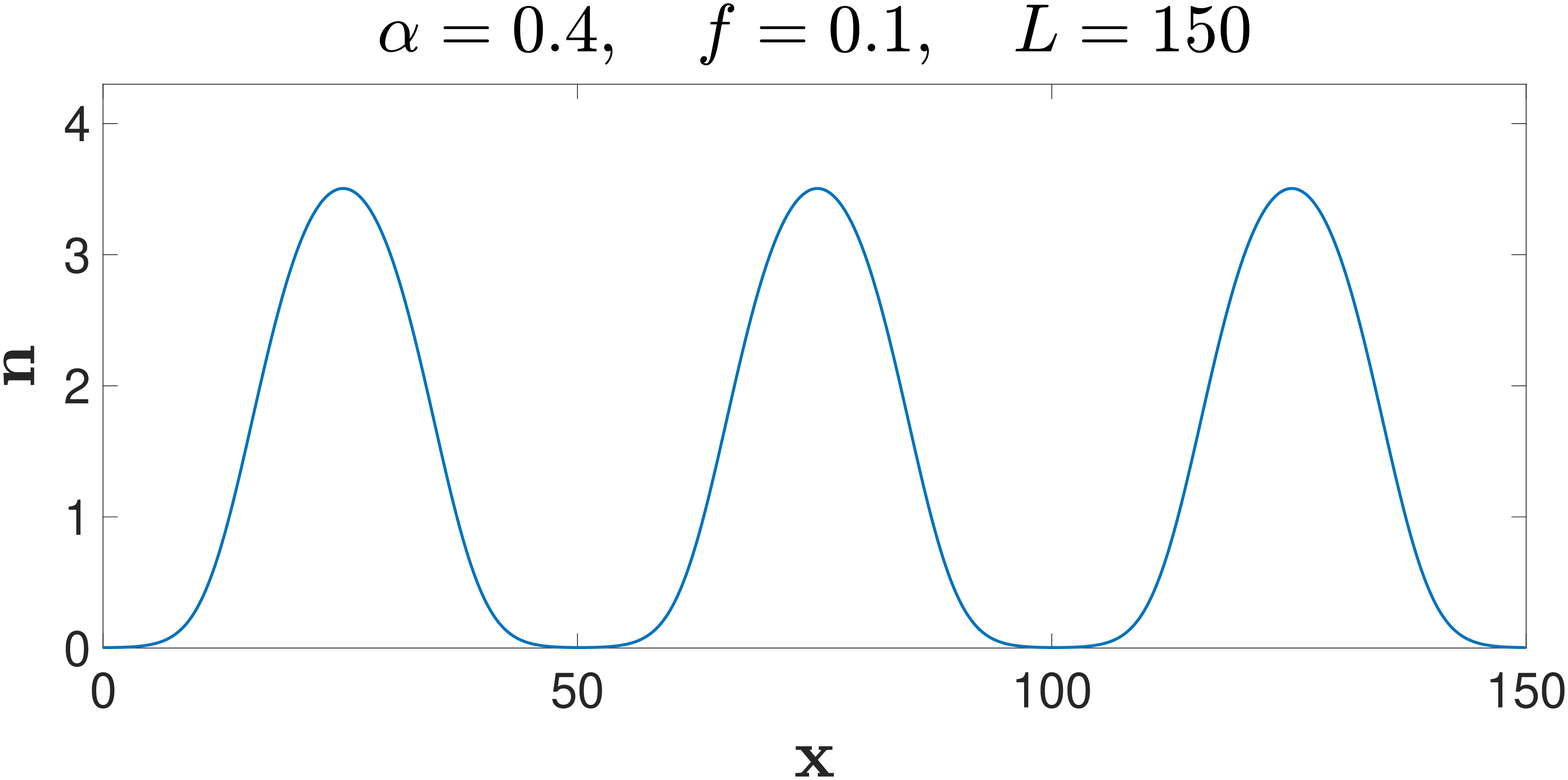}
        \caption*{(c) Red Branch}
    \end{subfigure}
    \caption{Graphs of biomass at $p\approx 0.35$ for each of the patterned branches shown in Fig.~(\ref{fig:Fig6}a).}\label{fig:Fig7}
\end{figure}

\begin{figure}  
\centering
\begin{subfigure}[c]{0.45\textwidth}
        \includegraphics[width=\textwidth, keepaspectratio=true]{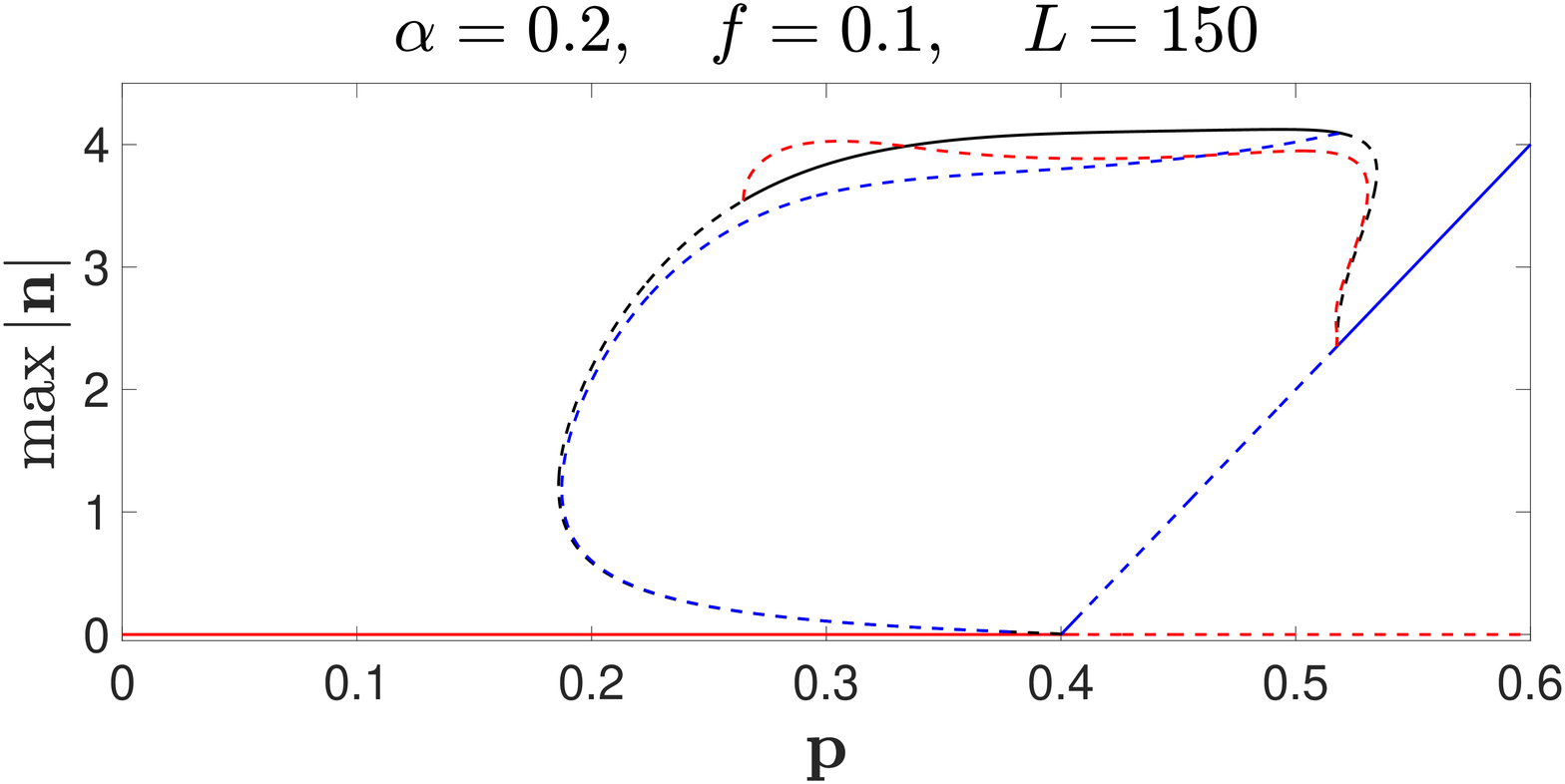}
        \caption*{(a) Bifurcation diagram showing the black branch from Fig.~(\ref{fig:Fig6}b) and secondary bifurcations. Note that the black branch consists of two equally spaced bumps.}
    \end{subfigure}
    \quad
    \begin{subfigure}[c]{0.45\textwidth}
        \includegraphics[width=\textwidth, keepaspectratio=true]{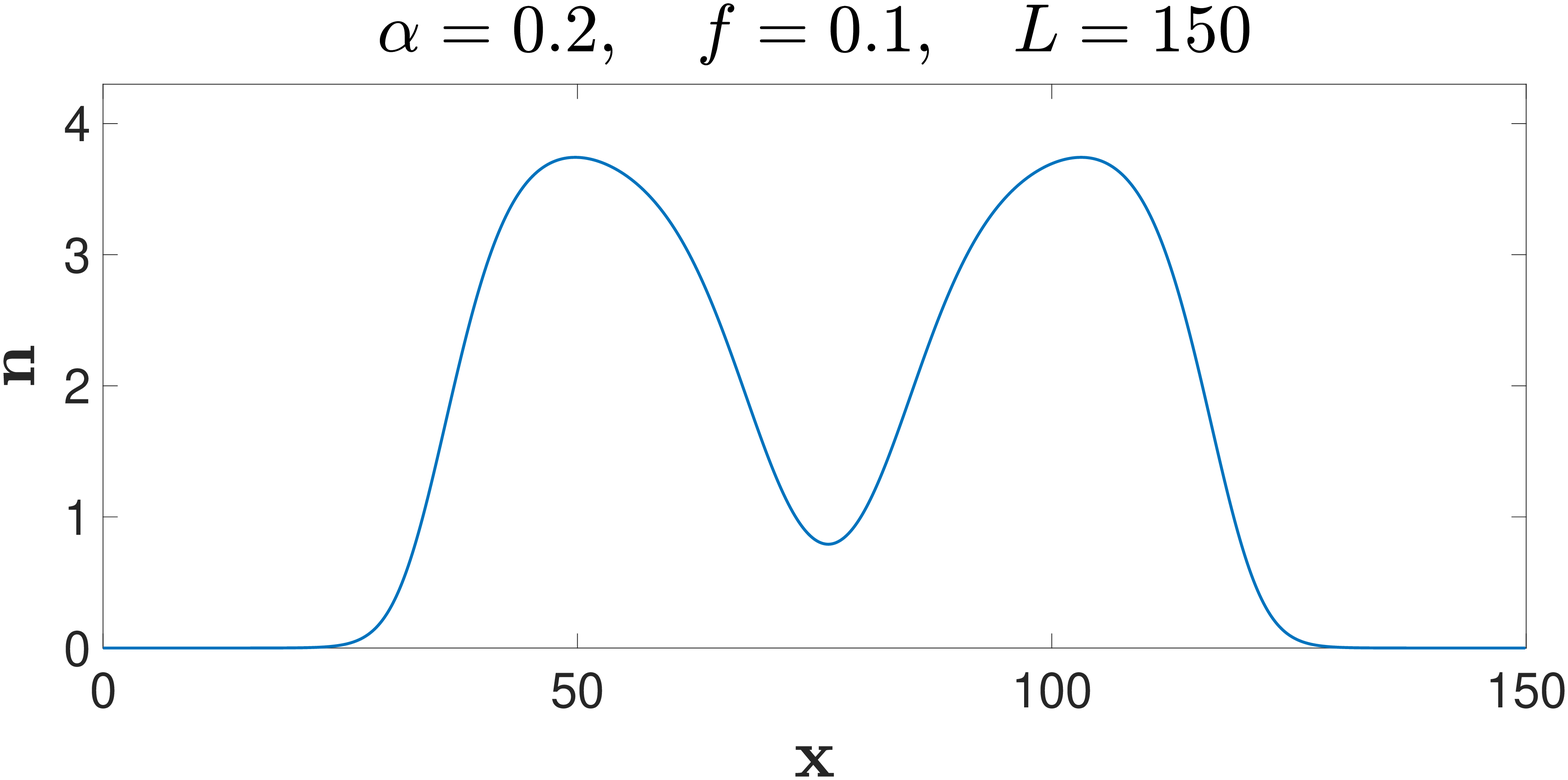}
        \caption*{(b) Here we see that the space between the bumps is unequal (in the sense that the space between them in the middle of the diagram versus the space between on the left and right of the diagram are not the same).}
    \end{subfigure}
    \quad
    \begin{subfigure}[c]{0.45\textwidth}
        \includegraphics[width=\textwidth, keepaspectratio=true]{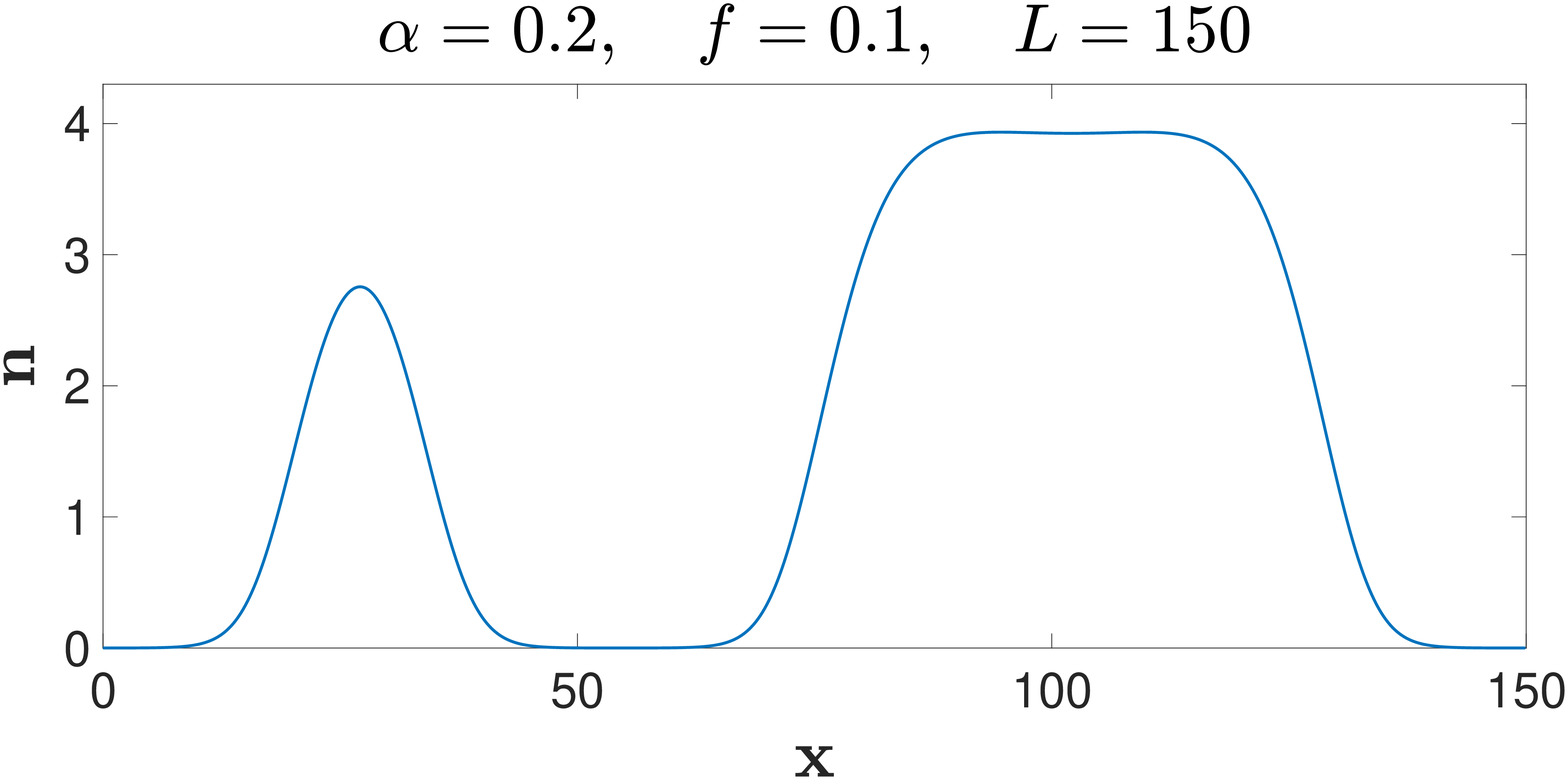}
        \caption*{(c) Here we see that the space between the bumps is equal, but that one bump is allowed to grow much larger than the other.}
    \end{subfigure}
    \caption{Symmetry breaking behavior arising from branch
      points along patterned branches and leading to states in panels
    (b) and (c) with either unequal space or size between them.}\label{fig:Fig8}
\end{figure}

We first consider Fig.~(\ref{fig:Fig6}a) and then discuss the others. Starting from the transcritical bifurcation at $p=0.4$ on the homogeneous vegetated branch, several pitchforks occur in quick succession as $p$ is increased. Amongst these, in the same location as for the corresponding $L=37$ diagram
(as explained above given its spatially independent character),
is a supercritical Hopf bifurcation. As expected, increasing
$p$ further, we additionally find several other pitchforks and a
subcritical Hopf bifurcation. We then  continue each of these patterned branches, three of which are shown in the diagram. Looking at the upper right corner, notice that all the patterned branches are initially unstable when they
bifurcate. This contrasts with what we found earlier on the smaller domain.
Furthermore, the largest branch extends rightward beyond its branch point, creating a region of bistability between a patterned inhomogeneous
and a vegetated homogeneous state. Near the middle of the branches, we see that two of the three become stable creating a region of multitability again
between different
patterned states. We also see that each of them has a Hopf bifurcation,
in line with what we observed earlier for the $L=37$ case.
Fig.~(\ref{fig:Fig7}a-c) shows the graph of the biomass for these branches at a selected point. We further remark that unlike with the $L=37$ case, we found several branch points along these patterned states. Fig.~(\ref{fig:Fig8}) a shows a bifurcation diagram, for $\alpha=0.2$ and $f=0.1$, which includes a continuation of two of these secondary bifurcations, while 
Fig.~(\ref{fig:Fig8}b) and Fig.~(\ref{fig:Fig8}c) show their symmetry breaking nature with potentially
either the shape or the size of the bumps becoming unequal in the
branches emerging from the patterned state.

The other diagrams all have a similar nature. We still see that as $\alpha$ increases the Hopf bifurcations on the patterned branches tend to approach the fold point. In Fig.~(\ref{fig:Fig6}d), we see a remarkable change in the largest branch, where it has apparently folded over on itself; its upper branch point also moves from $p \approx 0.5$ to $p \approx 0.4$ so that the branch still begins and ends at two different places on the vegetated branch. Perhaps the most remarkable feature is that the change from Fig.~(\ref{fig:Fig6}c) and Fig.~(\ref{fig:Fig6}d) is the result of a small change in one of the parameters. This suggests that qualitative results obtained in earlier works may also be quite sensitive to small variations in the parameters. Fig.~(\ref{fig:Fig6}e) is the companion to Fig.~(\ref{fig:Fig1}); we see, as before, that there are no Hopf bifurcations present. We also note that the maximum height and width of all the patterned branches decrease from Fig.~(\ref{fig:Fig6}d) to Fig.~(\ref{fig:Fig6}e), as would be expected since $f$ has increased.

\section{Unphysical Branches}

All previous studies to date have ignored bifurcations which occur on the
desert state for precipitation values beyond the transcritical
bifurcation or for values less than zero. In this section, we first
perform a Turing analysis to reveal the number and location of these bifurcations. We then continue some of these branches numerically and investigate their
behavior and the implications of their existence.

Recall that the desert state is given explicitly by
\[n_0=0, \quad w_0=\frac{p}{\nu}, \quad h_0=\frac{p}{f \alpha}.\]
We consider the linearization ansatz 
\[u_j(t) = {u_j}_0 + A_j e^{\lambda t}e^{ikx}\]
where the subscript $j=1, 2, 3$ runs over the 3 dependent
variables with $u_1=n$, $u_2=w$ and $u_3=h$. Substituting this
into Eq.~(\ref{eq1}) and linearizing, we arrive at the dispersion relationship 
\[0 = \det \bigg(\lambda I - DF(u_0) + \tilde{D}k^2 \bigg)\]
where in this case the Jacobian
\[DF(0,\frac{p}{\nu},\frac{p}{f \alpha})=\left(
\begin{array}{ccc}
-\mu + \frac{p}{p+\nu} \quad &  0  & \quad 0\\[.5cm]
\frac{(1-f)p}{f} - \gamma \frac{p}{p + \nu} & -\nu  & \alpha f\\[.5cm]
 -\frac{(1-f)p}{f} &  0   &-\alpha f\\
\end{array}
\right),
\]
is greatly simplified. Expanding, we have
\begin{eqnarray}
  \quad \lambda =  - \mu + \frac{p}{p+ \nu} - k^2 \quad , \quad -\nu - D_w k^2 \quad , \quad -f - D_h k^2.
  \end{eqnarray}
Since the latter two terms are always negative for realistic
(i.e., positive) parameter values, these
can never lead to branch points. Hence we only need to consider
\[\lambda(p,k) = - \mu + \frac{p}{p+ \nu} - k^2\]
For fixed $k$, a branch point occurs wherever $\lambda$ changes sign. Hence, setting $\lambda$ to zero and solving for $p$ gives
\[p = \frac{\nu(\mu + k^2)}{1 - \mu - k^2}.\]
This shows that the mode with wave number $k$ becomes unstable along the desert state at $p= \frac{\nu(\mu + k^2)}{1 - \mu - k^2}$.

It is straightforward to show that $p$ is an increasing function of $k^2$. We also have the following:

\[p(0) = \frac{\nu \mu}{1-\mu} \quad, \quad \lim\limits_{k^2 \to (1-\mu)^-} p = \infty \quad, \quad \lim\limits_{k^2 \to (1-\mu)^+} p=  -\infty \quad, \quad \lim\limits_{k^2 \to \infty} p = -\nu\]
Coupled with the fact that $p$ is increasing, the first two items in the list imply that $p([0,1-\mu)) =[\frac{\nu \mu}{1-\mu}, \infty)$ and the latter two items imply that $p((1-\mu,\infty)) = (-\infty,-\nu)$. Now, for any finite spatial interval $[0,L]$, only a discrete (infinite) set of wavenumbers are supported. If we order these wavenumbers in ascending order
\[k_0^2 , k_1^2 ,k_2^2  , \cdots\]
and allow $k_m^2 < 1-\mu < k_{m+1}^2$, then we have that 
\[p(\{k_0^2 , k_1^2 , \cdots, k_m^2 \}) \subset [\frac{\nu \mu}{1-\mu}, \infty) \quad , \quad  p(\{k_{m+1}^2 , k_{m+2}^2 , \cdots \}) \subset (-\infty,-\nu)\]
This last line immediately implies that there is a finite number of bifurcations to the right of the transcritical and infinitely many accumulating at $p=-\nu$.

\begin{figure}
\centering
    \begin{subfigure}[c]{0.45\textwidth}
        \includegraphics[width=\textwidth, keepaspectratio=true]{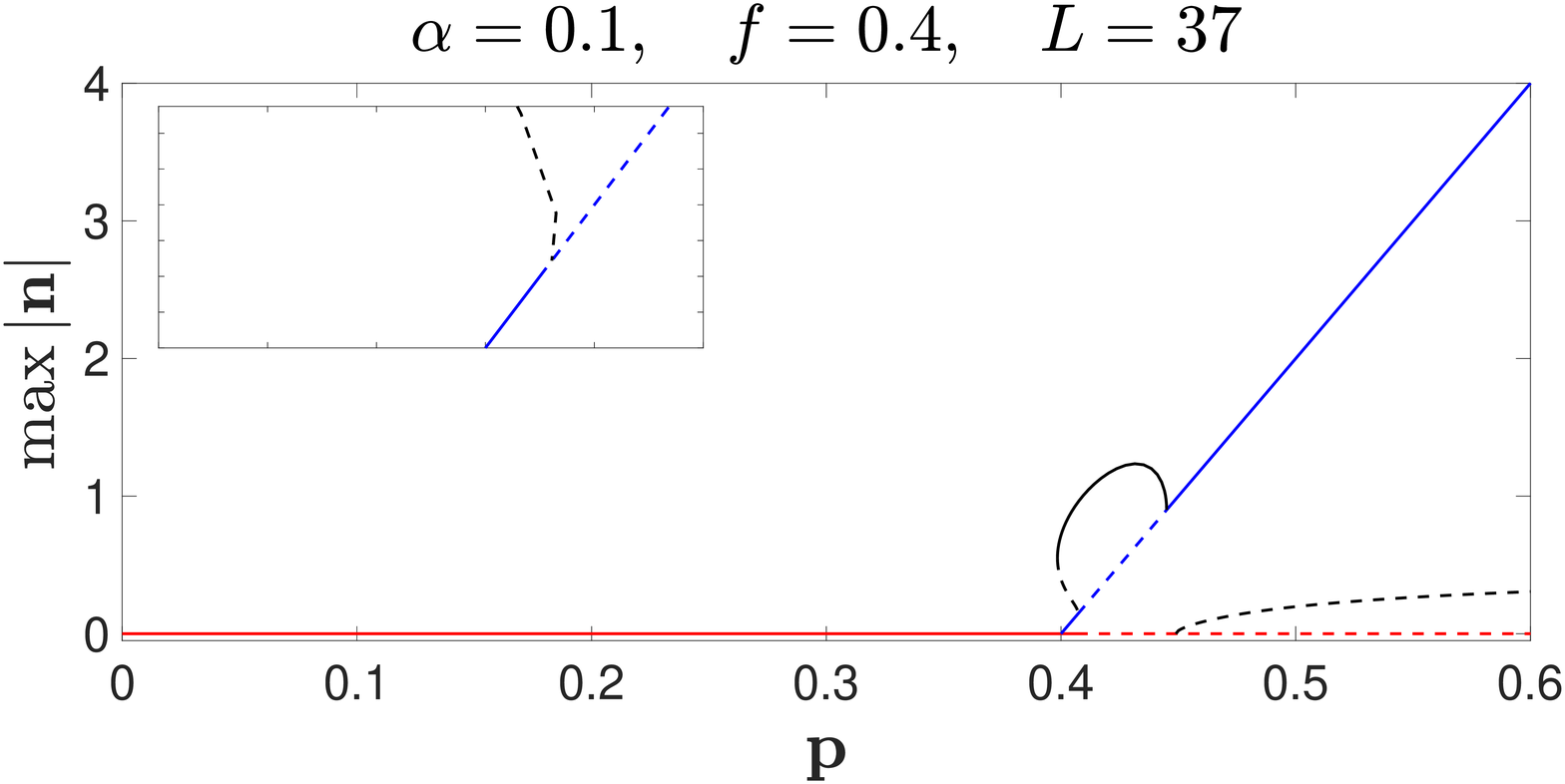}
        \caption*{(a)}
    \end{subfigure}
    \qquad
    \begin{subfigure}[c]{0.45\textwidth}
        \includegraphics[width=\textwidth, keepaspectratio=true]{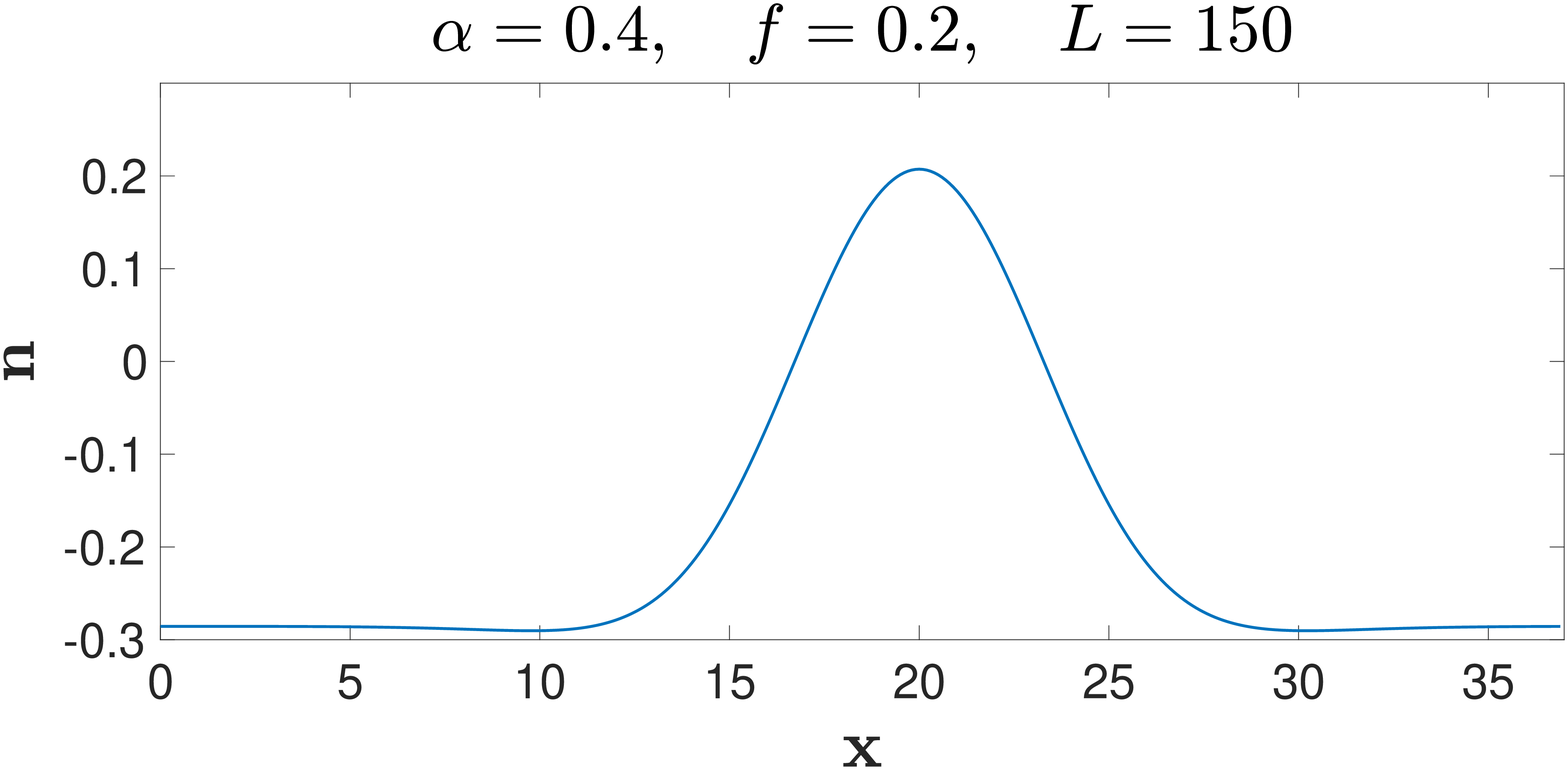}
        \caption*{(b)}
    \end{subfigure}
        \qquad
    \begin{subfigure}[c]{0.45\textwidth}
        \includegraphics[width=\textwidth, keepaspectratio=true]{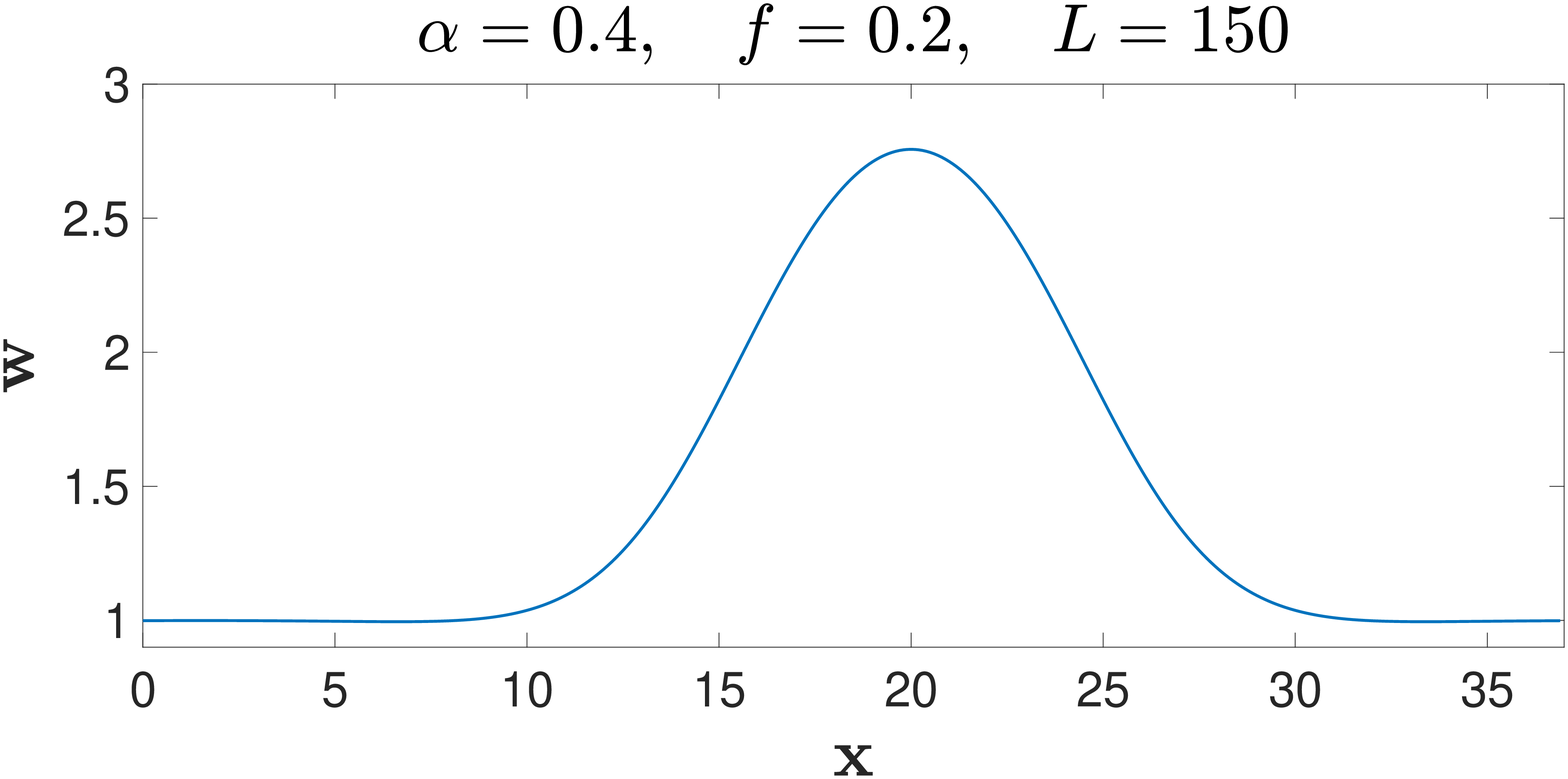}
        \caption*{(c)}
    \end{subfigure}
        \qquad
    \begin{subfigure}[c]{0.45\textwidth}
        \includegraphics[width=\textwidth, keepaspectratio=true]{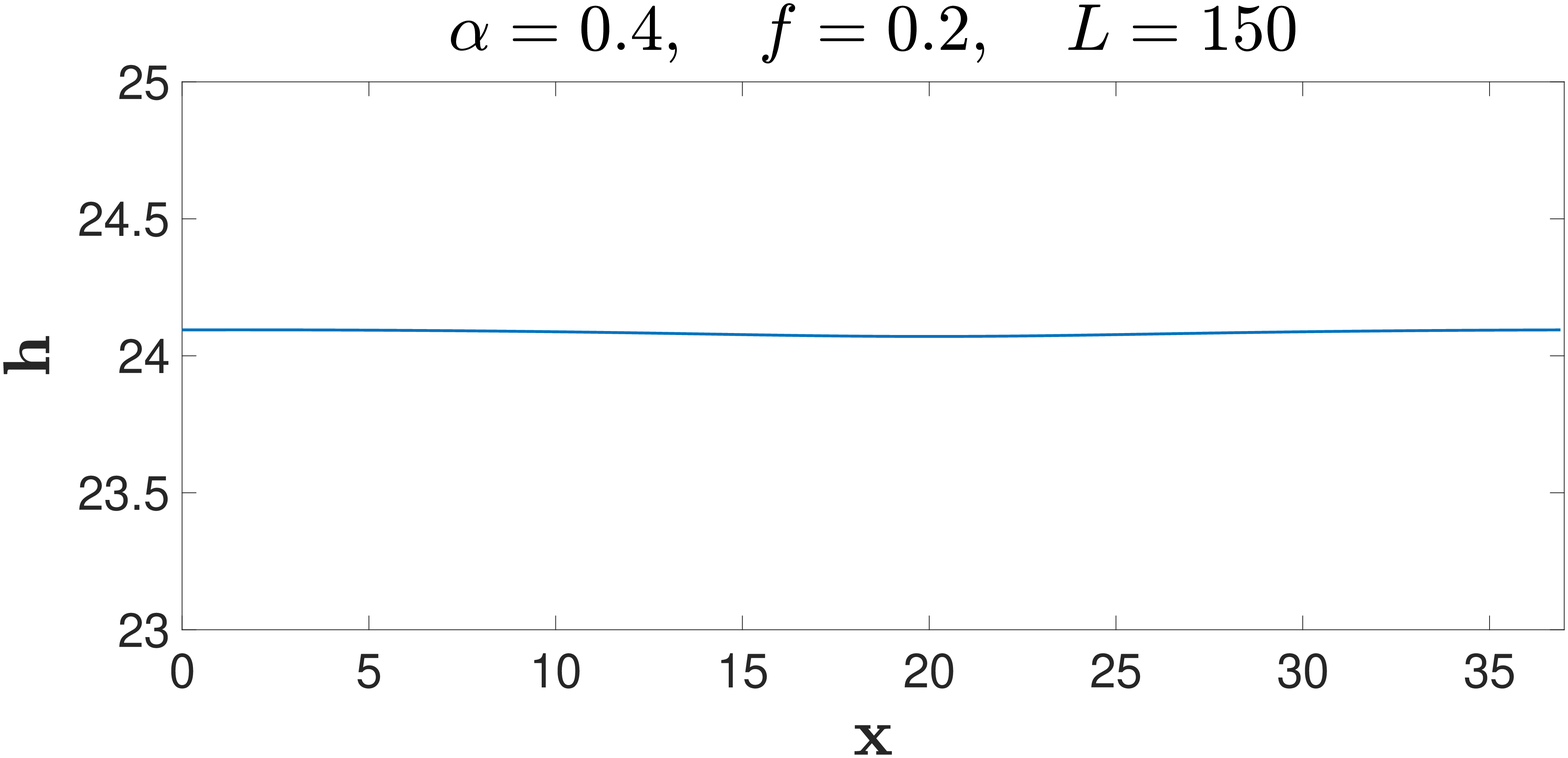}
        \caption*{(d)}
    \end{subfigure}
    \caption{Unphysical branch not shown in Fig.~(\ref{fig:Fig3}d).
      Notice that biomass takes negative values in (b).
      Shown also in (c) and (d) are the soil and surface water
    spatial profiles.} \label{fig:Fig9}
    
\end{figure}

Fig.~(\ref{fig:Fig9}a) shows the bifurcation diagram of Fig.~(\ref{fig:Fig3}d) with one such unphysical solution included. Panels (\ref{fig:Fig9}b-d) are the corresponding graphs of the spatial profiles of $n$, $w$ and $h$ of
the steady state for a selected point. There are several other branches that bifurcate off the desert branch for larger values of $p$, as implied
by our Turing analysis. For the $L=150$ case, there are actually several unphysical branch points within the interval $0 \leq p \leq 0.6$. In fact, one can show that the number of unphysical branches to the right of the transcritical
bifurcation
increases with domain size (due to the increase of the corresponding number
of accessible wavenumbers), and that they get closer and closer to the transcritical bifurcation point. Of the unphysical branches to the right of the transcritical that we have observed, it has always been the biomass which takes on negative values; we have never seen the surface water or soil water become
unphysical. Nevertheless, this clearly points to a pathology of the model.

\section{Modified Model}
In this section, we discuss a class of models, all of which are modified versions of RM, which do not possess infinitely many bifurcations from their desert state. We also numerically explore the simplest of these models which, in a certain sense, is a direct generalization of the model of Klausmeier
to three species.

Consider the class of models taking the form 
\begin{align}
\D{t}{n} &= -\mu n + G(n,w) n + \D[2]{x}{n} \nonumber \\ 
\D{t}{w} &= -\nu w + \alpha \frac{n + f}{n +1} h - \gamma G(n,w) n + D_w \D[2]{x}{w} \\ 
\D{t}{h} &= p - \alpha \frac{n + f}{n +1} h + D_h \D[2]{x}{h}, \nonumber
\end{align}
where we let $G(n,w)$ be a general growth term depending on both biomass and soil water. It can be shown that this model has two steady states. A desert state identical to the one previously considered, $(n_1,w_1,h_1) = (0,\frac{p}{\nu},\frac{p}{\alpha f})$, and a vegetated state which solves the equations
\[G(n_2,w_2)=0 \quad , \quad n_2=\frac{p - \nu w_2}{\gamma \mu} \quad, \quad h_2 = \frac{p}{\alpha} \frac{n_2 + 1}{n_2 +f}.\]
Doing a Turing analysis on the desert state, as in the last section, we arrive at the dispersion relation
\[0 = \det \bigg(\lambda I - DF(u_0) + \tilde{D}k^2 \bigg),\]
where in this case the Jacobian takes the form
\[DF(0,\frac{p}{\nu},\frac{p}{f \alpha})=\left(
\begin{array}{ccc}
-\mu + G(0,\frac{p}{\nu}) \quad &  0  & \quad 0\\[.5cm]
\frac{(1-f)p}{f} - \gamma G(0,\frac{p}{\nu})  & -\nu  & \alpha f\\[.5cm]
 -\frac{(1-f)p}{f} &  0   &-\alpha f\\
\end{array}
\right).
\]

Considering once again only the relevant eigenvalue (whose
zero crossings will provide the bifurcation points), we obtain
\begin{equation}\label{eq:dis}
\lambda(p,k) = -\mu + G(0,\frac{p}{\nu})  - k^2
\end{equation}
Setting $\lambda$ to zero, we see that bifurcations occur for those
values of $p$ and $k$ such that
\begin{equation}\label{eq:dis2}
\mu + k^2 = G(0,\frac{p}{\nu}).
\end{equation}
Now, suppose the solution is periodic on the domain $[0,L]$. Then $k$ can only take on a discrete, yet unbounded, set of values. Hence, because the LHS of Eq.~\eqref{eq:dis2} is an increasing, unbounded function of $k$, we see that if $G(0,\frac{p}{\nu})$ is bounded above then there can be at most a finite number of solutions to Eq.~\eqref{eq:dis2}. We then see that this class of models has a finite, possibly zero, number of bifurcations on the desert state if and only if $G(0,\frac{p}{\nu})$ is bounded above for all $p$ for which it is defined.
Two broad classes of such $G$ are:
\begin{itemize}
\item[$\bullet$] $G(n,w)=f(n)h(w)$, where $f(0)=0$. This leads to models in which no bifurcations are present. Such an example is $G=n w$.

\item[$\bullet$] $G(n,w)=f(n)h(w) + I(w)$, where $f(0)=0$ and $I(w)$ is bounded above. This leads to models where bifurcations are present.
  A relevant example, for instance, is $G=\frac{w^2}{w^2 +1}$.
\end{itemize}

When Rietkerk and collaborators generalized in~\cite{19}
the model of Klausmeier~\cite{17},
they added a surface water equation and changed the growth term. As we have seen, this growth term $\frac{w}{w+1}$ is not bounded (near $w=-1$) and so leads to infinitely many bifurcations from the desert state, a feature that
is pathological in the present context. For the remainder of the paper we
let the growth term take the form $G(n,w)=nw$, which is the original growth
rate used in the model proposed by Klausmeier. We note that this growth rate falls in the first category; it has no bifurcations from the desert state.

Our system then takes the form
\begin{align}
\D{t}{n} &= -\mu n + wn^2 + \D[2]{x}{n} \nonumber \\ 
\D{t}{w} &= -\nu w + \alpha \frac{n + f}{n +1} h - \gamma wn^2 + D_w \D[2]{x}{w} \\ 
\D{t}{h} &= p - \alpha \frac{n + f}{n +1} h + D_h \D[2]{x}{h}. \nonumber
\end{align}
The vegetated state for this model is given explicitly by
\[n_\pm = \frac{\mu}{w_\pm}\quad , \quad w_\pm =\frac{1}{2}A \pm \sqrt{A^2-4B} \quad, \quad h_\pm = \frac{p}{\alpha} \frac{n_\pm + 1}{n_\pm +f},\]
where $A=\frac{p}{\nu}$ and $B=\frac{\gamma \mu^2}{\nu}$.

\begin{figure}  
\centering
    \begin{subfigure}[c]{0.45\textwidth}
        \includegraphics[width=\textwidth, keepaspectratio=true]{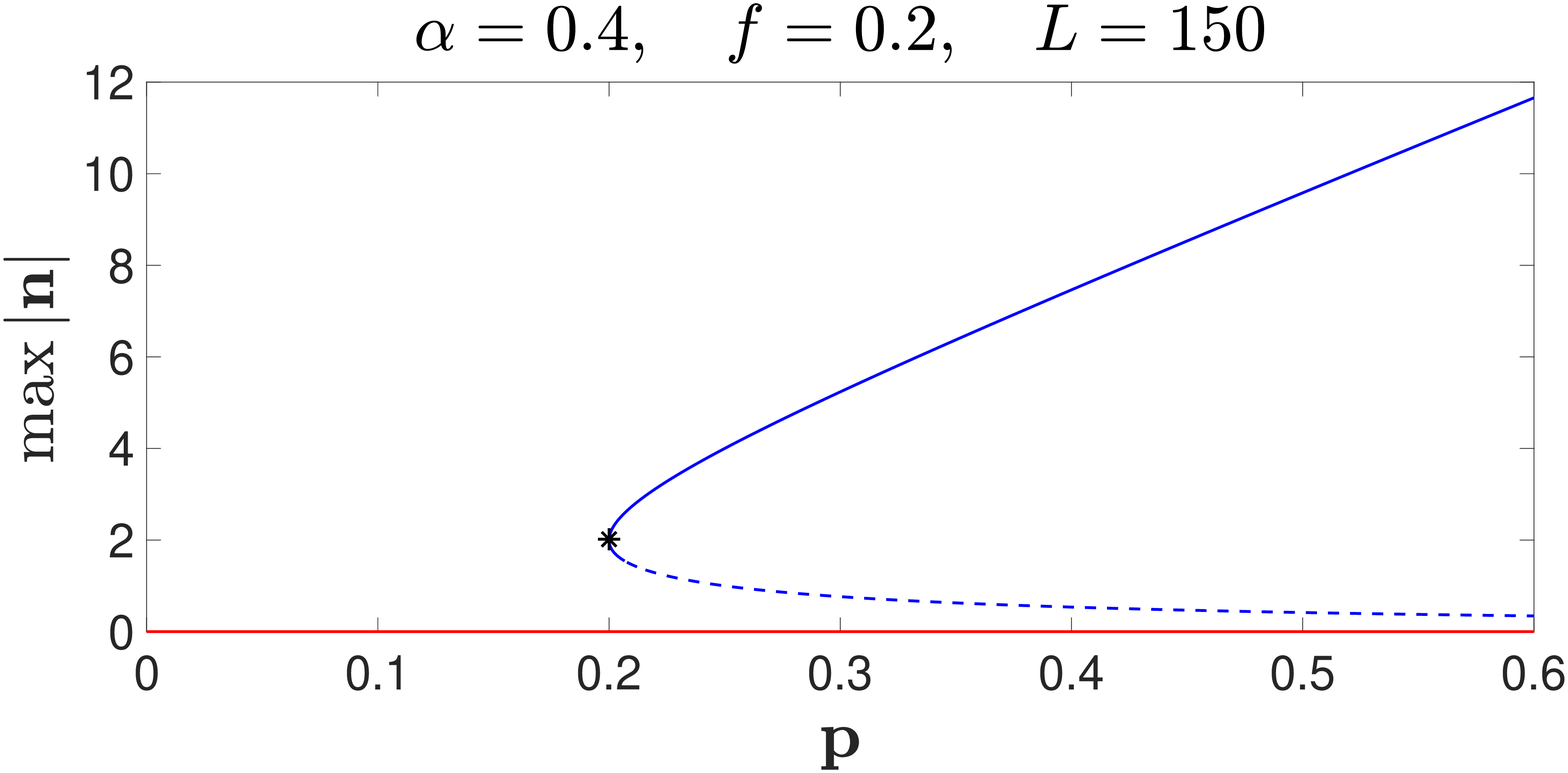}
        \caption*{(a)}
    \end{subfigure}
    \qquad
    \begin{subfigure}[c]{0.45\textwidth}
        \includegraphics[width=\textwidth, keepaspectratio=true]{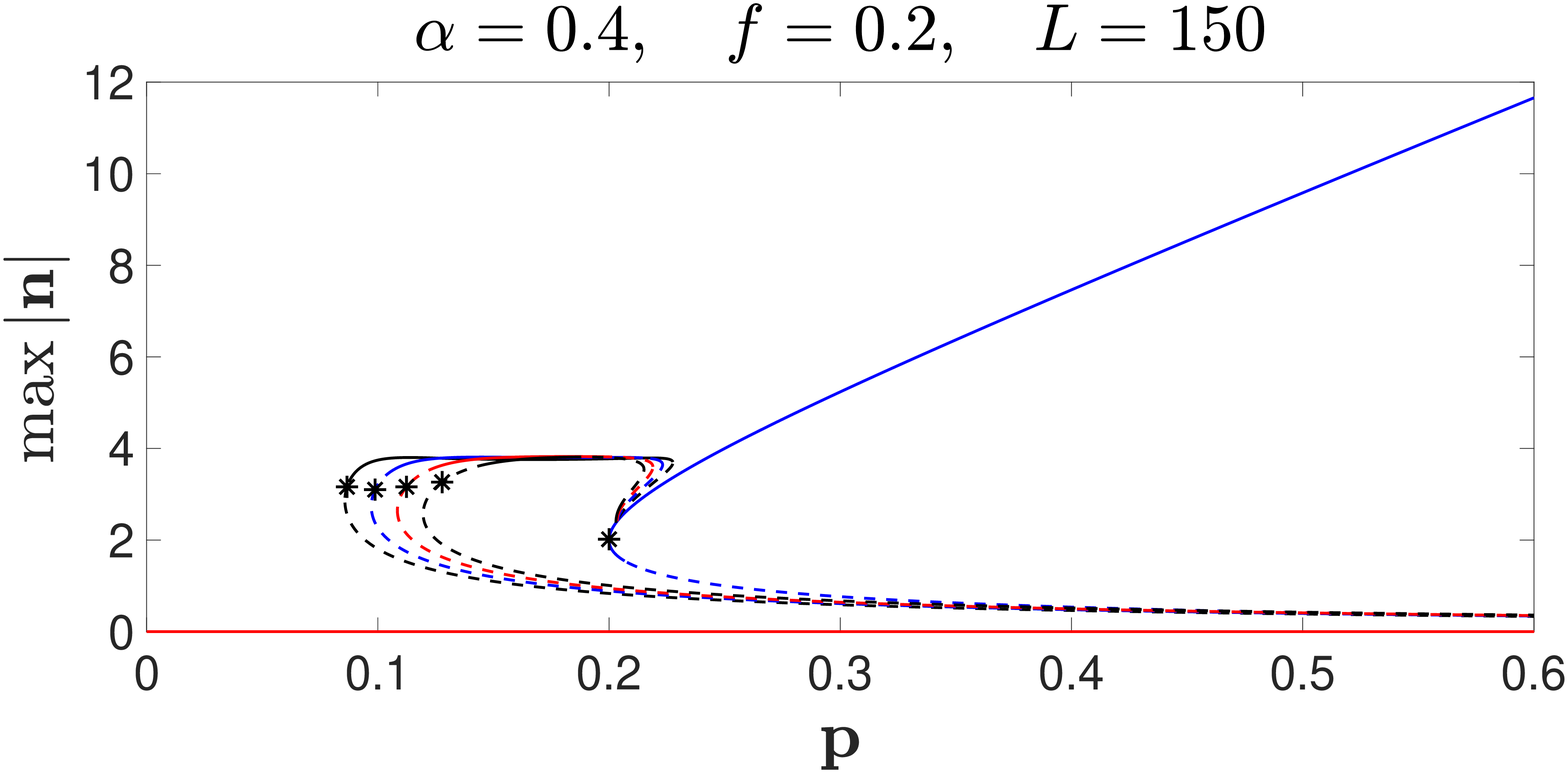}
        \caption*{(b)}
    \end{subfigure}
        \qquad
    \begin{subfigure}[c]{0.45\textwidth}
        \includegraphics[width=\textwidth, keepaspectratio=true]{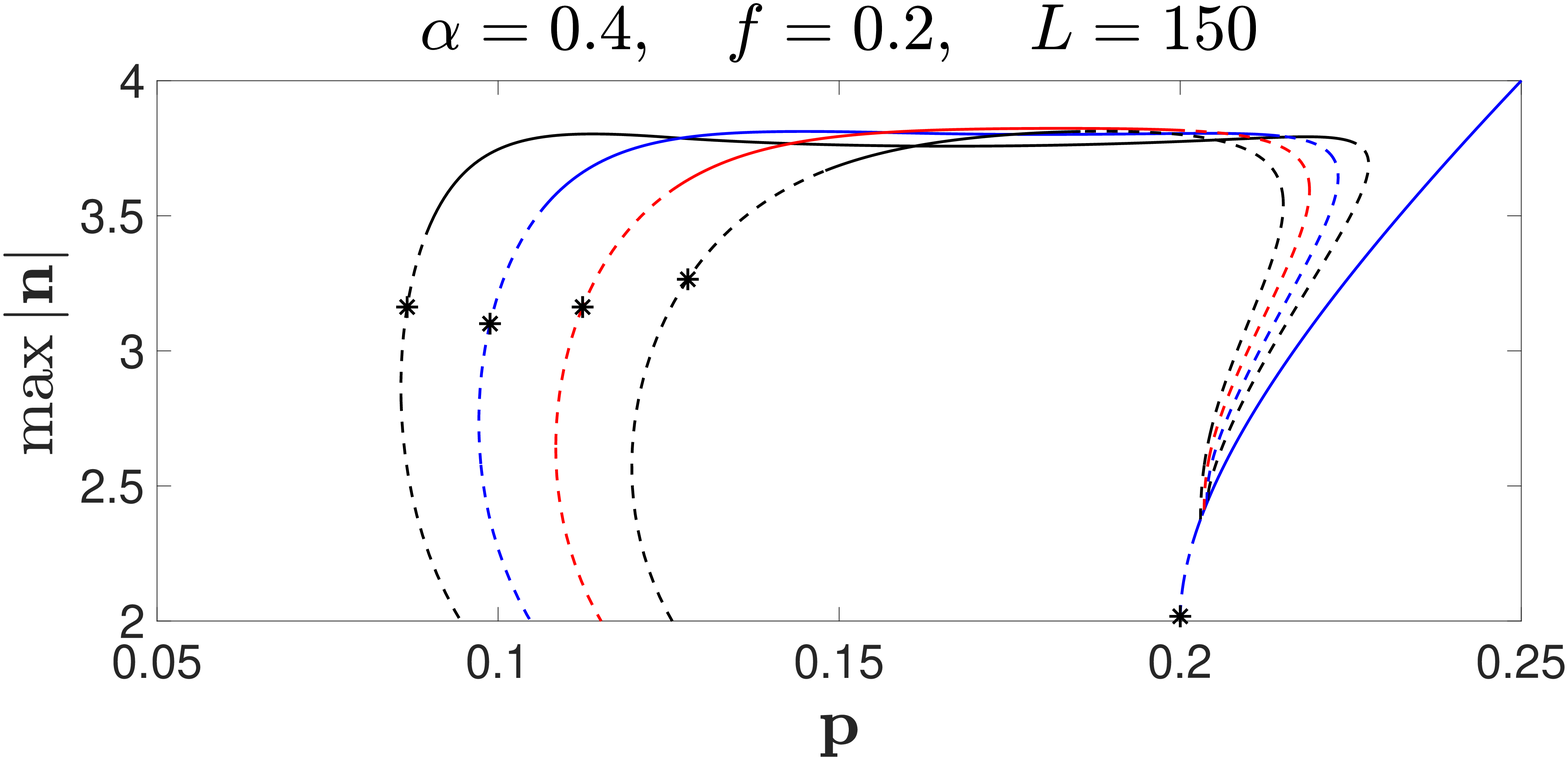}
        \caption*{(c)}
    \end{subfigure}
    \caption{Steady States in modified RM: relevant
      bifurcation diagrams are shown in the same variables as before
      (maximal biomass as a function of the precipitation parameter $p$).}
    \label{fig:Fig11}
\end{figure}

\begin{figure}  
\centering
    \begin{subfigure}[c]{0.45\textwidth}
        \includegraphics[width=\textwidth, keepaspectratio=true]{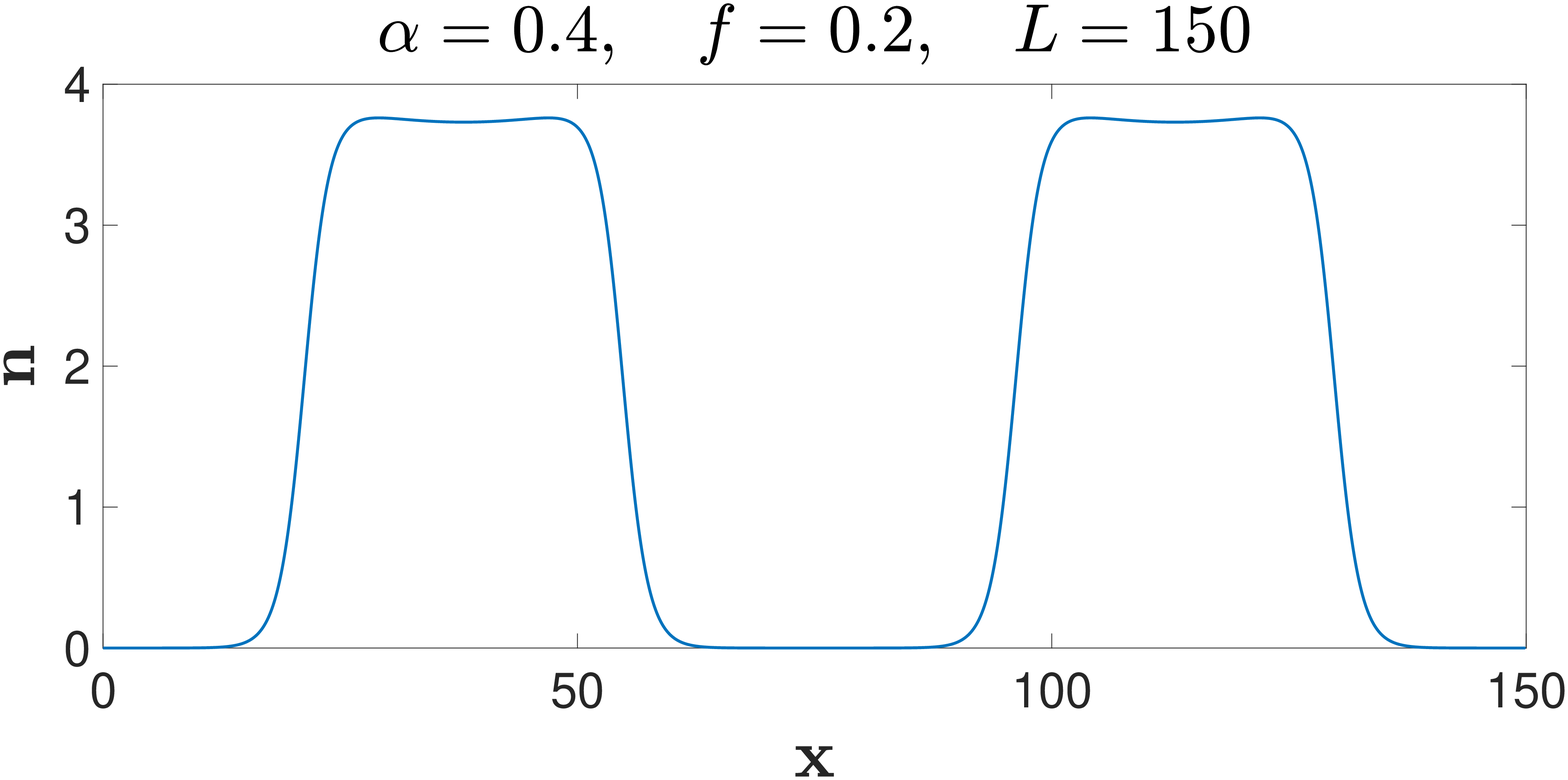}
        \caption*{(a)}
    \end{subfigure}
    \qquad
    \begin{subfigure}[c]{0.45\textwidth}
        \includegraphics[width=\textwidth, keepaspectratio=true]{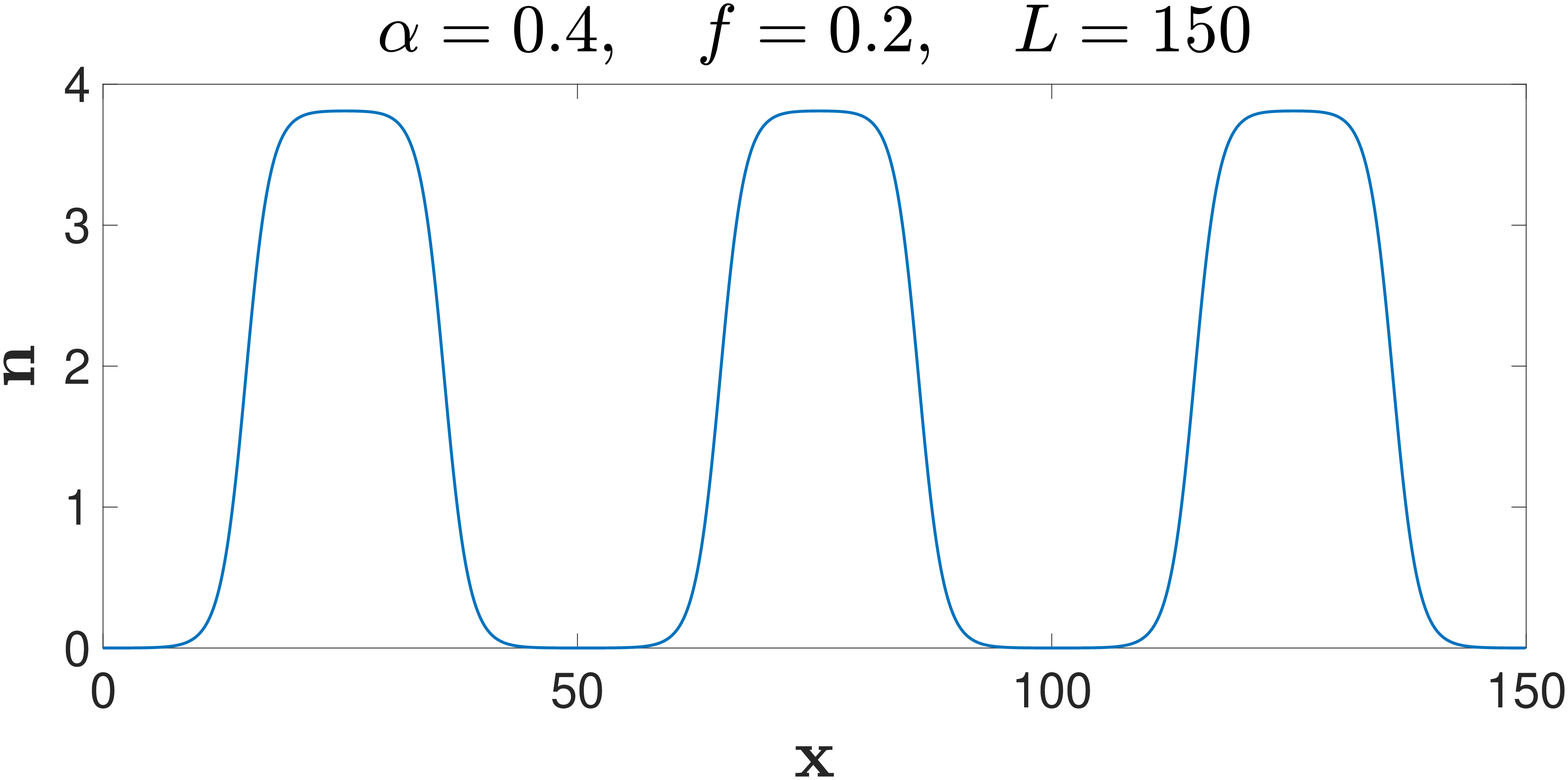}
        \caption*{(b)}
    \end{subfigure}
        \qquad
    \begin{subfigure}[c]{0.45\textwidth}
        \includegraphics[width=\textwidth, keepaspectratio=true]{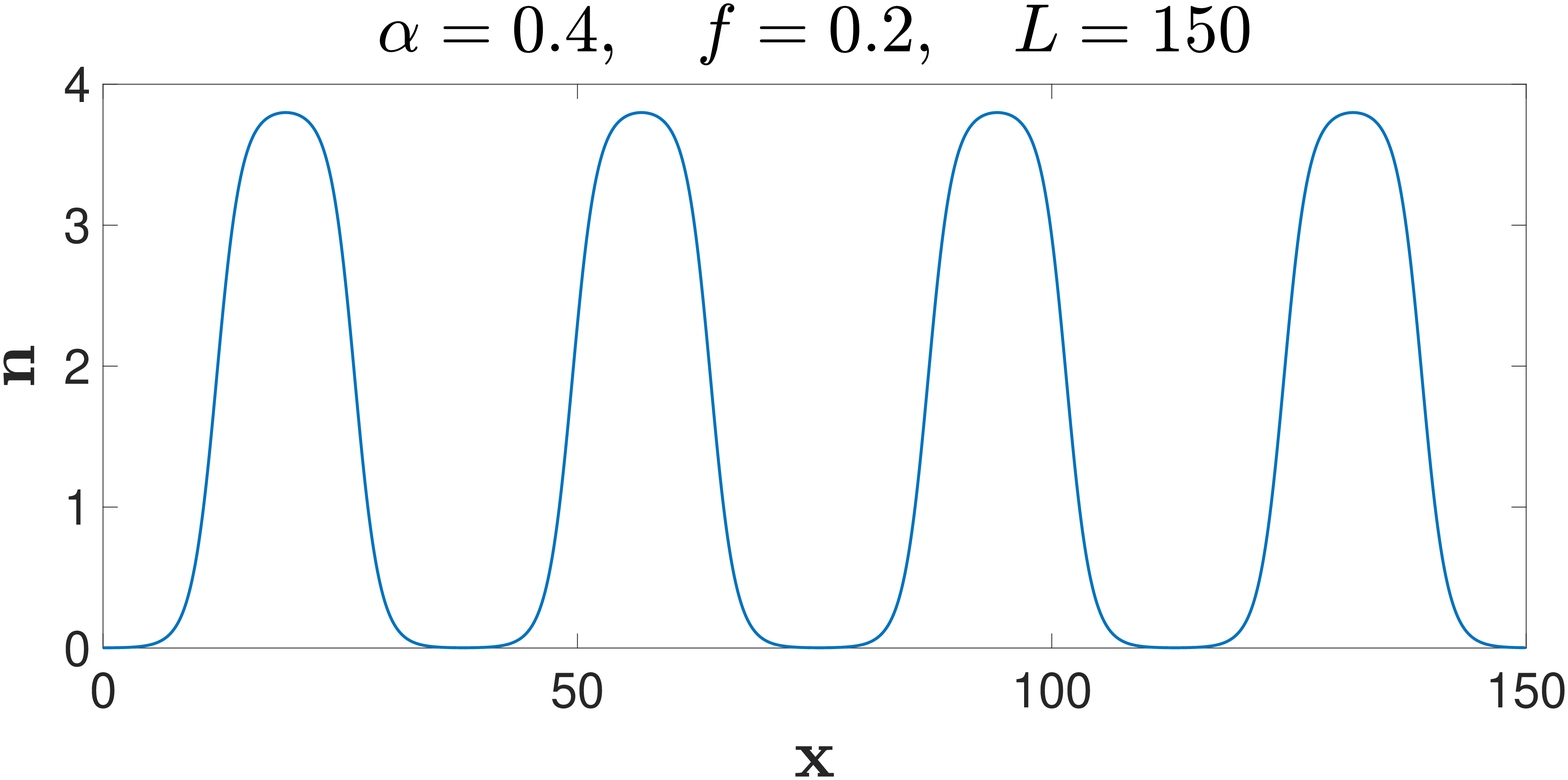}
        \caption*{(c)}
    \end{subfigure}
        \qquad
    \begin{subfigure}[c]{0.45\textwidth}
        \includegraphics[width=\textwidth, keepaspectratio=true]{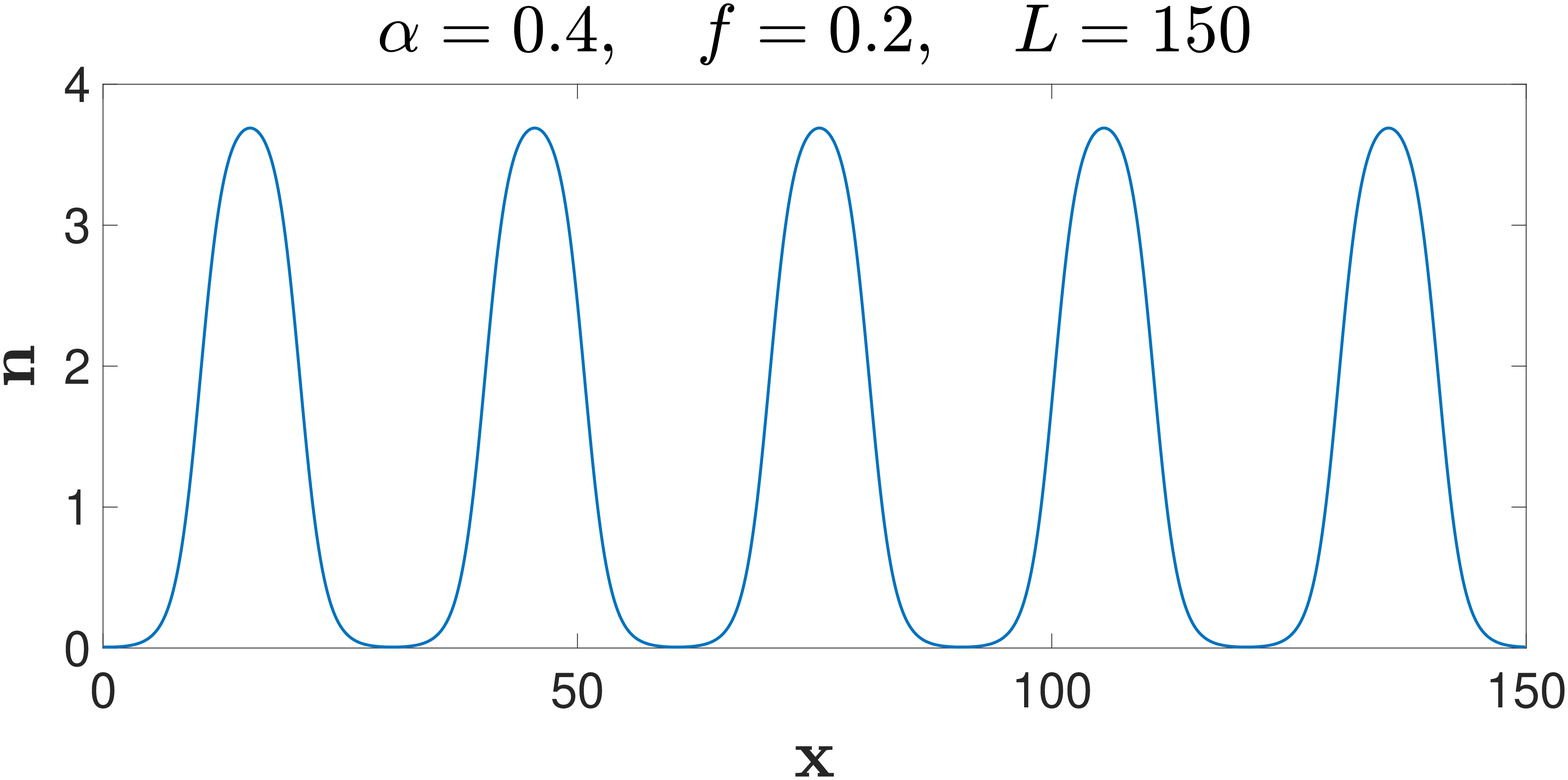}
        \caption*{(d)}
    \end{subfigure}
    \caption{Graph of biomass for each branch in the case
      of modified RM at $p \approx 0.15$.} \label{fig:Fig12}
\end{figure}

As in the previous sections, we discretize this system and
perform numerical continuation to construct the bifurcation diagrams of the steady states. For the diagrams that follow, we have chosen $\alpha=0.4$, $f=0.2$, and $L=150$ with the rest of the parameters fixed at the same values as before. Fig.~(\ref{fig:Fig11}a) shows a bifurcation diagram of the homogeneous steady states. As expected, the vegetated state does not bifurcate out of the desert state.
Instead, two such vegetated states terminate at a saddle-node
bifurcation when $A^2=4 B$ for the expressions above.
It's also worth mentioning that the desert state is stable to both homogeneous and inhomogeneous perturbations for all values of $p$; Eq.~(\ref{eq:dis}) shows that $\lambda$ is always a negative real number. Continuation
of the vegetated state shows that several bifurcations, including a Hopf bifurcation, occur. Fig.~(\ref{fig:Fig11}b) shows a bifurcation diagram where we have continued four of these patterned branches, and Fig.~(\ref{fig:Fig11}c) is a zoomed in version. As far as we can tell, the patterned branches begin on the vegetated homogeneous branch but then asymptotically
approach the desert branch as $p$ increases.
Fig.~(\ref{fig:Fig12}) shows a graph of the biomass at a given value
of the parameters ($\alpha=0.4$, $f=0.2$, $p=0.15$) for the different
branches of patterned steady state solutions.
We also note that the multistability range of the patterned states occurs for smaller values of precipitation than for the RM.
Finally, each of these patterned branches has a Hopf bifurcation, as
they are approaching their respective turning points, as can be seen,
e.g., in Fig.~(\ref{fig:Fig11}).

\section{Future Work and Discussion}  

In this work, we attempted to provide a systematic view of the bifurcation
diagram of the Rietkerk model for dryland vegetation, exploring the
possible states of the system. We varied both parameters of the
equation, such as the precipitation, and the parameters governing
the conversion of surface water to soil water; in parallel, we also
varied characteristics of the domain such as its size. We identified
numerically, and wherever possible analytically Hopf and pitchfork
points, converged on the periodic and asymmetric vegetated orbits
resulting from these bifurcations and considered the dynamics of
related instabilities. The potential of the system for hysteresis and
multi-stability was accordingly revealed. Multi-parameter bifurcation
diagrams were used to develop a sense of how the bifurcations are
modified under parametric variations. Pathologies of the model
(such as the bifurcation of an infinity of branches from the state
without vegetation) were uncovered, and a potential modification
of the model to avoid this effect was proposed and examined.

\begin{figure}
\centering
        \includegraphics[width=.45\textwidth, keepaspectratio=true]{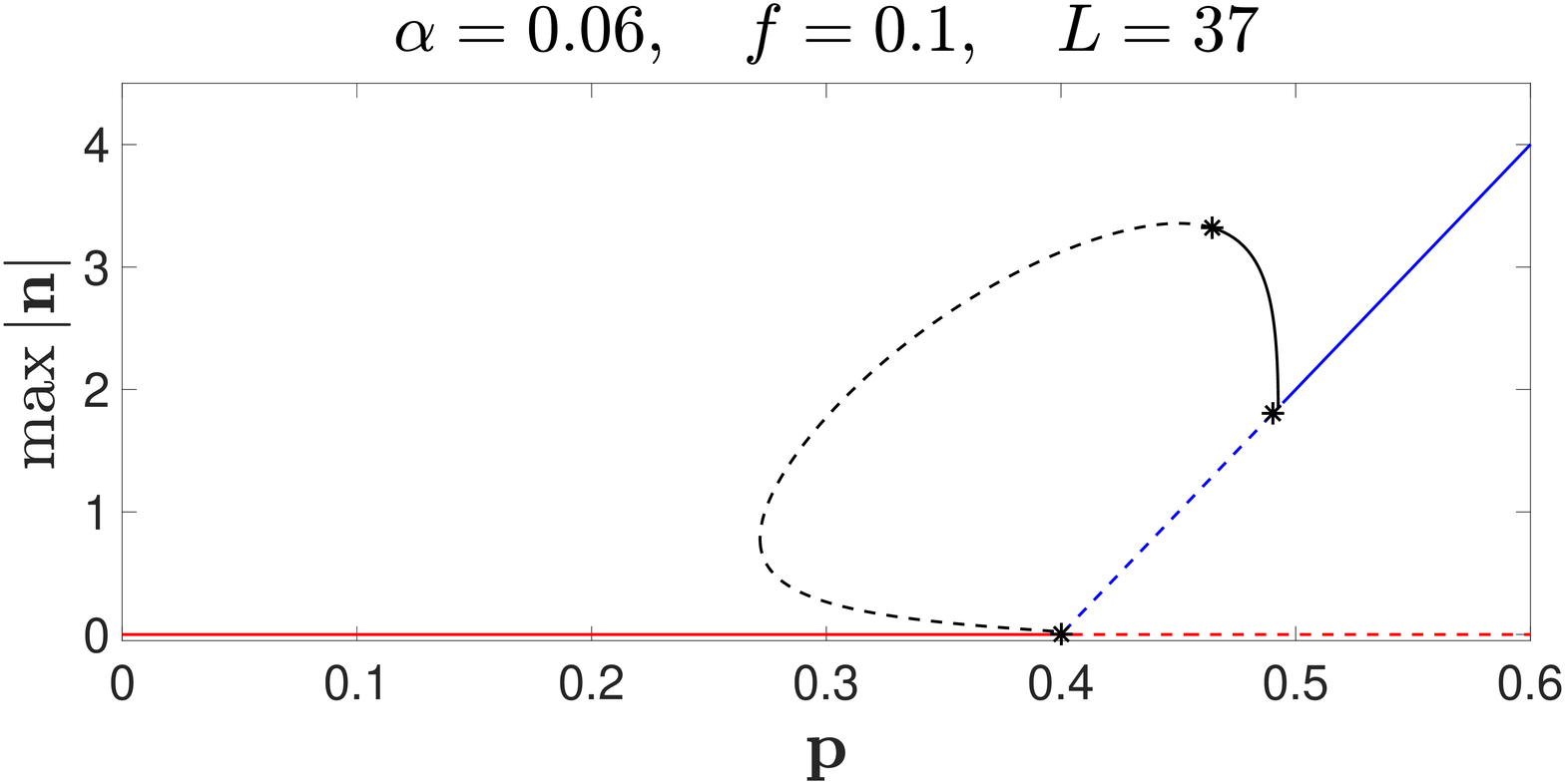}
        \caption{Bifurcation diagram for $\alpha=0.06$,
          $f=0.1$ and $L=37$, for the maximal biomass as a function
        of the precipitation rate $p$.}\label{fig:Fig13}
    \end{figure}

Several avenues for further work exist. Recall that in the second section we stated that around $\alpha=0.06$, the Hopf bifurcation on the patterned branch actually moves to the right of $p=0.4$. This can be seen in Fig.~(\ref{fig:Fig13}). Although we were unable to continue the homogeneous periodic orbit in this case, we integrated the system with the same procedure as before. In regions where the patterned state is unstable, we found the system always evolved to the periodic orbit, regardless of the perturbation. On the other hand, in regions where the patterned state was stable, we found that it generically evolved to the patterned state. It would be interesting to see how the stability of the periodic orbits actually change with the infiltration parameters. Furthermore, if for some parameter values the periodic orbits and the other steady states are all unstable, then it would
be especially intriguing to understand towards which state the system evolves.
More broadly, the detailed
investigation of these periodic orbits in 1D could be a particularly
interesting topic in its own right.

Further work on the bifurcation diagrams for larger domain sizes would also be of interest. The choice of $L=150$ led to three branches which had regions of stability. On larger domains more patterned states would be stable and it would
become relevant to acquire a sense of
how small changes in $\alpha$ and $f$ affect the shape of these patterned
branches.
Understanding also the origin and extent of parametric relevance
of the folding over behavior would also constitute a relevant question.

There is also the natural generalization of these results to 2D which
warrants further exploration.
In that context, several studies have been devoted to predicting how an ecosystem transitions from patterned to desert states; see,
e.g., the discussion of the recent Ref.~\cite{4} and references therein.
A common finding is that they transition via the "gaps $\rightarrow$ labyrinth $\rightarrow$ spots" sequence.
Understanding the role (to such transitions) of the wide parametric variations
considered here and the impact of features such as hysteresis and
multi-stability would be particularly relevant to examine in that
context.

Last but certainly not least, we indicated the existence of infinitely many unphysical branches in the model. This is a feature that seems to us that
it would be preferable
to generally avoid. 
In a different direction, although we did not remark on it above, similarly
to the desert branch, there may also exist
infinitely many branch points on the vegetated state at $p=\frac{\nu \mu}{1 - \mu}-\gamma \mu$. It is at this $p$-value that the biomass assumes
the value negative one; recalling that the infiltration rate is $\alpha \frac{n + f}{n+1}h$, we see that the system is singular here as well.
This leads us to wonder if the infiltration term could be modified,
in a way similar to what was done
for the growth term, in such a way that these bifurcations do not occur and
otherwise yielding results in qualitative agreement with RM. 

Recalling the class of modified RMs that we proposed, there is much work to
be done in the exploration and evaluation of these. We chose to explore the
case of $G(n,w)=nw$ model specifically because it has the simplest form and because it constitutes a direct generalization of the Klausmeier model
to three species. We note that this model also eliminates
the bifurcations at $p=\frac{\nu \mu}{1 - \mu}-\gamma \mu$. The appearance of Hopf bifurcations in this new model suggests to us that
these may be a generic feature; specifically, since they depended on
$\alpha$ and $f$ in such a strong way, we suspect that it is
the infiltration term which gives rise to these.
Exploring some of these directions will be deferred to future
studies.\\

\textbf{Acknowledgments} The authors are particularly grateful to Prof. M. Silber and Dr. S. Iams 
for numerous insightful discussions on the topic and for many crucial remarks
towards shaping the present work.

\section{Appendix}

Here, we provide some details about the numerical and analytical
calculations presented througout the manuscript.

\subsection{Numerical Methods}
We study steady state solutions of Eq.~(\ref{eq1}) in the form of a boundary value problem on $[0,L]$:
\begin{equation}\label{eq2}
0 =F(u) + \tilde{D}  \Do[2]{x}{u}  \\[-.2cm]
\end{equation}
\begin{equation}\label{eq2b}
u(0) = u(L), 
\end{equation}
where 
\[u=
\left( n,w,h\begin{array}{cc}
\end{array}
\right)^T\]

\[ F(u)=
\left( \begin{array}{c}
-\mu n + \frac{w}{w+1} n \\[.3cm]
-\nu w + \alpha \frac{n + f}{n +1} h - \gamma \frac{w}{w+1} n\\[.3cm]
p - \alpha \frac{n + f}{n +1} h
\end{array} \right)
\]
\\
\[
\tilde{D}=\left( \begin{array}{ccc}
1 & 0 & 0\\
0 & D_w & 0\\
0 & 0 & D_h 
\end{array}
\right)
\]
and Eq.~\eqref{eq2b} denotes \emph{periodic boundary conditions}. Now, due to invariance under spatial translations, to make the BVP well defined a standard trick is to add a constraint equation which uniquely picks out one of the translates \cite{7}. One such choice is to demand that the derivative of the biomass be zero at some fixed point in the domain. While, at least formally, this makes the BVP well-defined, any discretization of it will automatically have one more equation than unknowns; this means a direct application of Newtons method would be impossible. To get around this, we consider instead the traveling wave equation with periodic boundary conditions and a slightly modified constraint equation:
\begin{align}\label{eq3}
&0 =F(u) + c\frac{du}{dx} + \tilde{D}  \Do[2]{x}{u} \nonumber \\
&u(0) = u(L)   \\
&n'(0)+c=0. \nonumber
\end{align}
Formally, this BVP is well-defined. Further, after discretizing, the number of equations will equal the number of unknowns, allowing Newtons method to be used. It also important to note that any solution of this BVP which has $c=0$ is automatically a solution of Eq.~(\ref{eq2}).

We discretize Eq.~(\ref{eq3}) with $\Delta x = \frac{L}{N}$, choosing $N$ so that $\Delta x = 0.1$. Spatial derivatives are approximated by second-order center difference formulas with the periodic boundary conditions incorporated. This results in a large number of algebraic equations which one can then solve for using Newton's method: 

\begin{align}\label{eq4}
0&=F(u_1) + c \frac{u_2 - u_N}{2 \Delta x}+\tilde{D}  \frac{u_{2} - 2u_{1} + u_{N}}{\Delta x ^2}  \nonumber \\[.3cm] 
0&=F(u_i) + c \frac{u_{i+1} - u_{i-1}}{2 \Delta x} +\tilde{D}  \frac{u_{i+1} - 2u_{i} + u_{i-1}}{\Delta x ^2} \nonumber \\[.3cm] 
0&=F(u_{\raisebox{-1.5pt}{$\scriptstyle N$}}) + c \frac{u_1 - u_{N-1}}{2 \Delta x} + \tilde{D}  \frac{u_{1} - 2u_{N} + u_{N-1}}{\Delta x ^2}  \\[.3cm] 
0&=\frac{n_1 - n_{N}}{2 \Delta x} + c  \nonumber 
\end{align}
where \small $1< i <N$. We henceforth denote this system by $G(u,p)=0$.

Since we are primarily interested in how steady states change as the precipitation changes, we do not just want to solve one BVP but a family of such problems. To do this, one typically guesses an initial solution $u_0$ and then uses Newton's Method to obtain a true solution. One then increases a parameter and uses the previously obtained solution as the new initial guess. The process is then repeated until some stop condition is met. A more sophisticated way of doing parameter continuation is via pseudo-arclength continuation \cite{7,8}.  The idea here is to impose a constraint equation demanding that the system be parametrized by arclength. It has several advantages over standard parameter continuation such as being able to pass through fold points and the simple detection of (simple) branch points.

There are many standard continuation packages which incorporate numerical bifurcation methods \cite{9,10,11}. For ODE boundary value problems and some simple PDE boundary value problems, these programs tend to work quite well. However, for more complex PDE boundary value problems, such as in multiple dimensions or with many symmetries present, these packages may be insufficient. In our case, the biggest inconvenience is that they can not detect multiple branch points. While there exist methods for detecting multiple branch points these tend to be computationally expensive \cite{12}. Since any branch point has to occur where the jacobian of Eq.~(\ref{eq4}) becomes singular, the simplest solution to finding branch points, in which an odd number of branches emanate outward, is to look for changes in the sign of the determinant of the jacobian, $\det DG$; this is used in the software pde2path, for example. However, this does not detect branch points in which an even number of branches emanates outward, and so it is of limited use.

An alternative to this is to look at the ratio 
\begin{equation}\label{logdet}
\frac{d}{dp}[\log(|\det DG(p)|)]=\frac{(\det DG(p))'}{\det DG(p)}
\end{equation}
(where we have implicitly assumed $u=u(p)$ along the branch). If the jacobian becomes singular at an isolated point (and assuming $\det DG(p)$ is locally analytic at this point) then we see that not all of the derivatives of $\det DG$ can vanish (else it wouldn't be isolated). Expanding $\det DG$ in Taylor series (and setting to zero only a finite number of the coeffecients) we see that the RHS behaves like $1/p$. Hence, around isolated singularities we see that this ratio changes sign, and will thus find all possible branch points (again, assuming the function is locally analytic).

Depending on the implementation, the LHS of Eq.~\eqref{logdet} can be calculated from the LU decompositions of the jacobian of the current iterate and the previous iterate. Hence the cost is minimal. Once a branch point is detected along the branch, we use this function in a Newton method to find the precise location of the branch point. To be specific, a Newton step for finding 
the zeros of $\det DG(p)$ takes the form
\[p_{k+1} = p_k - \frac{\det DG(p)}{(\det DG(p))'}\]

As an additional check for branch points and for the detection of Hopf bifurcations, we simply calculate all the eigenvalues of the jacobian after the continuation has completed. In the cases where a Hopf bifurcation is detected, we use MATCONT to do the continuation of the periodic orbit.

\subsection{Turing Analysis on Vegetated Branch}
Consider again  Eq.~(\ref{eq2}). Given any steady-state $u_0$, we can expand the equations about it using the ansantz
\[u(t) = u_0 + Ae^{\lambda t}e^{ikx}.\]
Plugging this into (\ref{eq2}) and linearizing gives
\begin{IEEEeqnarray}{rCl}
\lambda Ae^{\lambda t}e^{ikx} &=& \bigg[DF(u_0)- \tilde{D}k^2 \bigg]Ae^{\lambda t}e^{ikx} \nonumber \\[.2cm]
\Rightarrow \quad  \quad \lambda A &=& \bigg[DF(u_0)- \tilde{D}k^2 \bigg]A  \nonumber  \\[.2cm]
\Rightarrow \quad \quad \quad 0 &=& \det \bigg(\lambda I - DF(u_0) + \tilde{D}k^2 \bigg) \label{eigeq}
\end{IEEEeqnarray}
where $DF(u)=DF(n,w,h)$ is given by

\[DF(n,w,h)=\left(
\begin{array}{ccc}
-\mu + \frac{w}{w+1} \quad &\frac{n}{(w+1)^2}& \quad 0\\[.5cm]
\alpha \frac{1-f}{(n+1)^2}h - \gamma \frac{w}{w+1}& -\nu - \gamma \frac{n}{(w+1)^2}&\alpha \frac{n+f}{n+1}\\[.5cm]
-\alpha \frac{1-f}{(n+1)^2}h&0&-\alpha \frac{n+f}{n+1}\\
\end{array}
\right).
\]
\\
\noindent
Eq.~(\ref{eigeq}) gives the relationship between $\lambda$, $u_0$, and $k$. If $u_0$ depends on a bifurcation parameter $p$, then $\lambda$ becomes a function of both $p$ and $k$. For fixed $k$, we expect to see a bifurcation in the nonlinear system (\ref{eq2}) whenever the real part of $\lambda(p,k)$ changes sign. In particular, if a complex conjugate pair crosses through the real axis then we expect to see a Hopf Bifurcation in the system.\\

\noindent
Plugging Branch 2, given explicitly by
\[n_0 = \frac{1}{\gamma \mu} (p - \frac{\nu \mu}{1-\mu}) \quad , \quad w_0 = \frac{\mu}{1-\mu} \quad, \quad h_0 = \frac{p}{\alpha} \frac{n_0 + 1}{n_0 +f}\]
into Eq.~(\ref{eigeq}) and expanding gives
\begin{IEEEeqnarray}{rCl}
c(\lambda)&:=&\det \bigg(\lambda I - DF(u_0) + \tilde{D}k^2 \bigg) \nonumber \\[.2cm]
&=& (\lambda + k^2)(\lambda + \nu + \gamma(1-\mu)^2n_0 + k^2D_w)(\lambda + \frac{p}{h_0} + k^2D_h) \nonumber \\[.2cm]
&& + ((1-\mu)^2 n_0)\bigg[(-\alpha \frac{1-f}{(n_0+1)^2}h_0 + \gamma \mu)(\lambda + \frac{p}{h_0} + k^2D_h) \nonumber \\[.2cm]
&& - (\alpha \frac{1-f}{(n_0+1)^2}h_0 )(-\frac{p}{h_0})\bigg]. \label{eigeqbr2}
\end{IEEEeqnarray}
Setting $k=0$ in Eq.~(\ref{eigeqbr2}) and simplifying gives
\begin{IEEEeqnarray}{rCl}
c(\lambda)&=&\lambda^3 + \bigg(\nu + \gamma(1-\mu)^2 n_0 + \frac{p}{h_0}\bigg)\lambda^2 \nonumber \\[.2cm]
&&+ (1-\mu)^2 n_0 \bigg( \frac{\nu p}{(1-\mu)^2 n_0 h_0}+ \frac{\gamma p}{h_0}-\alpha \frac{1-f}{(n_0+1)^2}h_0 + \gamma \mu\bigg)\lambda \nonumber \\[.2cm]
&& + \bigg(\frac{(1-\mu)^2n_0 \gamma \mu p}{h_0}\bigg). \nonumber
\end{IEEEeqnarray}

\noindent
Now, a polynomial $\lambda^3 + a_2 \lambda^2 + a_1 \lambda + a_0$ has a pair of pure imaginary roots \cite{13} precisely when
\[
\begin{cases}
a_1 >0 &\\[.2cm]
a_1a_2-a_0 =0.
\end{cases}
\]

\noindent
Letting $C=\frac{\nu p}{(1-\mu)^2 n_0 h_0}+ \frac{\gamma p}{h_0}-\alpha \frac{1-f}{(n_0+1)^2}h_0$, we have that $c(\lambda)$ has a pair of purely imaginary roots iff

\begin{subnumcases}{}
C + \gamma \mu >0 & \label{HBcond} \\[.2cm]
\bigg(\nu + (1-\mu)^2n_0\gamma + \frac{p}{h_0}\bigg)C + \bigg(\nu \gamma \mu +(1-\mu)^2 n_0 \gamma^2 \mu\bigg) =0. \label{HBeq}
\end{subnumcases}

\noindent
This implies a Hopf bifurcation will occur on the vegetated branch precisely when Eqs.~(\ref{HBcond}), (\ref{HBeq}) hold. Solving these equations and doing continuation gives Fig.~(\ref{fig:Fig5}a) from earlier.

\end{document}